\documentclass[useAMS,usenatbib,referee]{biom}
\usepackage{adjustbox}
\usepackage{graphicx}
\usepackage{amsmath}
\usepackage{amssymb}
\usepackage{multirow}
\usepackage[tight,footnotesize]{subfigure}
\usepackage{hyperref}
\usepackage{algcompatible}
\usepackage{algorithm}
\usepackage[noend]{algpseudocode}
\usepackage{xcolor}

\makeatletter
\def\BState{\State\hskip-\ALG@thistlm}
\makeatother

\usepackage{amsmath}
\DeclareMathOperator*{\argmax}{arg\,max}
\DeclareMathOperator*{\argmin}{arg\,min}

\def\bSig\mathbf{\Sigma}

\title[Estimating the optimal linear combination of predictors using SCOR]{ \vskip-1cm Estimating the optimal linear combination of predictors using  spherically constrained optimization}

\author{Priyam Das$^{1,*}$\email{Priyam\_Das@hms.harvard.edu}, 
Debsurya De$^{2}$, Raju Maiti$^{3}$,
Mona Kamal$^{4}$, Katherine A. Hutcheson$^{5}$,\\
{\bf Clifton D.\ Fuller$^{\mathbf{4}}$, Bibhas Chakraborty$^{\mathbf{3,6,7}}$ and Christine B.\ Peterson$^{\mathbf{8}}$} \\
$^{1}$Department of Biomedical Informatics, Harvard Medical School, USA \\
$^{2}$Indian Statistical Institute, Kolkata, India\\
$^{3}$Centre for Quantitative Medicine, Duke-National University of Singapore Medical School, Singapore\\
$^{4}$Department of Radiation Oncology, The University of Texas MD Anderson Cancer Center, USA \\
$^{5}$Department of Head and Neck Surgery, The University of Texas MD Anderson Cancer Center, USA \\
$^{6}$Department of Statistics and Applied Probability, National University of Singapore, Singapore\\
$^{7}$Department of Biostatistics and Bioinformatics, Duke University, USA\\
$^{8}$Department of Biostatistics, The University of Texas MD Anderson Cancer Center, USA}

\begin{document}


\date{{\it Received $\ldots$} 2020. {\it Revised $\ldots$} 2020.  {\it
Accepted $\ldots$} 2020.}



\pagerange{\pageref{firstpage}--\pageref{lastpage}} 
\volume{$\ldots$}
\pubyear{$\ldots$}
\artmonth{$\ldots$}


\doi{10.1111/j.1541-0420.2005.00454.x}


\label{firstpage}


\begin{abstract}
In the context of a binary classification problem, the optimal linear combination of continuous predictors
can be estimated by maximizing an empirical estimate of the area under the receiver operating characteristic (ROC) curve (AUC).
For multi-category responses, the optimal predictor combination can similarly be obtained by maximization of the empirical hypervolume under the manifold (HUM).
This problem is particularly relevant to medical research, where it may be of interest to diagnose a disease with various subtypes or predict a multi-category outcome.
Since the empirical HUM is discontinuous, non-differentiable, and possibly multi-modal, solving this maximization problem
requires a global optimization technique.
Estimation of the optimal coefficient vector 
using existing global optimization techniques is computationally expensive, becoming prohibitive as the number of predictors and the number of outcome categories increases. 
We propose an efficient derivative-free black-box optimization technique based on pattern search to solve this problem.
Through extensive simulation studies, we demonstrate that the proposed method achieves better performance
compared to existing methods including the step-down algorithm. Finally, we illustrate the proposed method to predict swallowing difficulty after radiation therapy for oropharyngeal cancer based on radiation dose to various structures in the head and neck.\\ \vskip-.15cm
\end{abstract}

%



\maketitle


%

\section{Introduction}
\label{sec1}
In precision medicine, there is great interest in utilizing multiple biomarkers or continuous variables to categorize patients based on disease subtype, predict treatment response, or anticipate the potential severity of toxicity. To address this challenge, various classification methods have been proposed. Existing model-based approaches to the classification problem include logistic regression and linear discriminant analysis. As an alternative to model-based methods, \cite{Pepe2000} and \cite{Pepe2006}  focused on classifying different subgroups based on maximization of the area under the receiver operating characteristic (ROC) curve (AUC).
In its simplest form, the ROC curve can be interpreted as the probability of detection as a function of the false alarm rate, so higher values of the AUC signify higher average hit rates over different possible values of the false alarm rate. In the field of medical science, the AUC is widely used as a criteria to combine multiple diagnostic test results or other continuous predictor values \citep{ Zhou2002}. 
However, the ROC curve does not have an exact analytical expression, so estimating the AUC and finding optimal predictor combinations to maximize its value remains challenging.

In the case of a binary outcome, various non-parametric estimates of the AUC have been proposed. For example, \cite{Pepe2006} estimated the AUC using Mann-Whitney U statistics, while \cite{Ma2007} proposed a smooth estimate obtained using a sigmoid function. However, with increasing sample size, these objective functions become challenging to maximize. In order to handle this problem for large datasets, \cite{Liu2011} proposed the min-max method, which only considers the linear combination of two extreme biomarkers in estimating the AUC, and therefore maintains the same computation time for any given sample size.
For multi-categorical outcomes, the 
hypervolume under the manifold (HUM), also known as the volume under the ROC surface, has been proposed as an analogue to the AUC  \citep{Li2008}. For the three-category outcome scenario, various estimates of the HUM have been proposed \citep{Scurfield1996, Mossman1999, Nakas2004}. 
In the multi-category setting, \cite{Yanyu2010} proposed a method to estimate the optimal combination vector when the outcomes of the biomarkers follow a normal distribution. However their proposed method was noted to under-perform when the biomarker values are non-normal \citep{Hsu2016, Maiti2019}. In the presence of multiple biomarkers, to avoid the computational burden of simultaneous estimation of the combination vector, \cite{Pepe2006} proposed the step-down algorithm. However, this strategy of maximizing the HUM estimate by estimating one coordinate at a time can work poorly in higher dimensional settings.
More recently, \cite{Zhang2011} proposed an empirical estimate of HUM (EHUM) for the three-category outcome scenario, but because of the lack of concavity of most of the proposed estimates, maximization of the objective functions remained challenging. To overcome this computational burden, under the assumption of normality, \cite{Kang2013} proposed a penalized and scaled stochastic distance based method. Finally, \cite{Hsu2016} proposed an efficient approach which relies on upper and lower bounds (ULBA).


As discussed above, most of the proposed estimates of the HUM are discontinuous and non-differentiable functions of the combination coefficients, and potentially have multiple maximums. Therefore,
one of the most vital aspects regarding their performance remains the maximization step. This problem becomes more challenging with increasing sample size, number of biomarkers, and the number of outcome categories. For simple cases, even in the presence of multiple local maximums, most of the optimization algorithms used in practice can still find the best solution.
But, as the maximization problem gets harder, it becomes more difficult to find the global maximum out of all local maximums. Thus,  performance becomes highly dependent on the optimization algorithm used for the maximization. When maximizing discontinuous, non-differentiable, multi-modal objective functions, derivative-free non-convex optimization techniques are critical to ensure a good solution \citep{Horst2002}. However, to the best of our knowledge, none of the previous articles proposing HUM estimates considered using global optimization techniques for maximizing the corresponding objective function. Our goal in the current work is both to point out the limitation of existing methods for this problem, and to propose a global optimization method that offers improved  performance in the prediction of multi-category outcomes.

\subsection*{Global optimization using pattern search}
We now give a more precise mathematical description of the problem of interest.
Suppose $\boldsymbol{\beta}$ denotes the $d$-dimensional linear combination vector to be estimated, and $f(\boldsymbol{\beta})$ denotes the empirical value of the HUM 
The way this objective functions is defined, if $f(\boldsymbol{\beta})$ is optimized over the unconstrained space $\boldsymbol{\beta} \in \mathrm{R}^d$, $f(\boldsymbol{\beta})$ is not identifiable as $f(\boldsymbol{\beta}) = f(a\boldsymbol{\beta})$ holds true for any given $a>0$ \citep{Liu2011}. In order to address this non-identifiability, instead of maximizing $f(\boldsymbol{\beta})$ over $\mathrm{R}^d$, we add an extra constraint $||\boldsymbol{\beta}|| = 1$ where $||\cdot||$ denotes the Euclidean norm. So the problem can be re-defined as
\begin{align}
maximize:& \; f(\boldsymbol{\beta}), \text{ where }\boldsymbol{\beta} = (\beta_1,\ldots,\beta_d) \nonumber \\
subject \; to:& \; \sum_{i=1}^d\beta^2_i = 1, \boldsymbol{\beta} \in \mathrm{R}^d,
\label{opt_problem}
\end{align}
where $f(\boldsymbol{\beta})$ can be discontinuous (as is the case for EHUM), non-differentiable, and multi-modal. Note that here the coordinates $\boldsymbol{\beta}$ must lie on the surface of a unit sphere. Due to  the possibly multi-modal nature of $f(\boldsymbol{\beta})$,  global optimization tools are preferred over convex optimization methods. 

There are a number of existing optimization methods, which we summarize briefly here. 
For maximizing any multi-modal function, global optimization techniques such as the genetic algorithm \citep{Fraser1957} and simulated annealing \citep{Kirkpatrick1983} have been shown to yield better results as compared to convex optimization methods such as the interior-point algorithm \citep{Karmakar1984} or sequential quadratic programming \citep{Boggs1996}.
The Nelder-Mead algorithm \citep{Nelder1965}, also known as the simplex method, is a heuristic derivative-free optimization technique that has been widely applied to statistical problems, and serves as the default option for the \texttt{optim} function in R and the basis of \texttt{fminsearch} in MATLAB. To improve on the convergence of these algorithms,
\cite{Torczon1997} proposed pattern search (PS), where possible solution points around the current solution are found using an adaptive step-size vector. Importantly, none of these  algorithms were specifically designed to handle global optimization over a spherical parameter space.

To minimize a non-convex function (or, equivalently, maximize a non-concave function) globally over a hyper-rectangular parameter space, \cite{Das2016} proposed a modified version of global pattern search called Recursive Modified Pattern Search (RMPS). This approach, which is discussed in detail in Section \ref{sec2_basic}, has desirable properties in terms of computational scalability and the fact that many of the operations can be performed in parallel. Subsequently, \cite{Das2016b} and \cite{Das2020} extended RMPS to problems where the parameter space is given by a collection of simplexes.

In this article we develop an algorithm called `Spherically Constrained Optimization Routine' (SCOR) which utilizes the basic principle of the RMPS algorithm but accommodates the constraint of a spherical parameter space. Based on an extensive set of simulation studies, the computational efficiency of the proposed algorithm is established and its performance is compared with existing optimization techniques. We show that using SCOR to maximize EHUM or ULBA results in noticeable improvement over the Nelder-Mead, step-down and min-max algorithms. Finally, we apply our proposed algorithm to derive a combination score which best predicts the severity of dysphagia, or difficulty swallowing, based on the mean radiation dose delivered to various structures in the head and neck during cancer treatment.

The rest of the article is organized as follows. Section \ref{sec_bkgd} summarizes various measures of the HUM and existing techniques for finding the optimal combination coefficients. Section \ref{sec_algo} describes the proposed SCOR algorithm and its theoretical properties. In Section \ref{sec_sim}, the performance of the proposed algorithm algorithm is compared to that of existing methods a series of challenging simulation studies. In Section \ref{sec_appl}, the SCOR algorithm is used to find the optimal combination coefficient vector to predict different grades of swallowing difficulty following radiation treatment for oropharyngeal cancer. Finally, we conclude with a discussion in Section \ref{sec_conc}.

\section{Background on estimating and optimizing the HUM}
\label{sec_bkgd}
In this section, we describe existing methods of estimating the optimal coefficient vector for combining biomarkers. Consider a study with $M$ possible outcome categories. Let $\mathbf{X}_1, \ldots, \mathbf{X}_M$ denote the $d$-dimensional observed biomarker values for the $M$ outcome categories. Each coordinate of $\mathbf{X}_j$ denotes the value of a biomarker, $j=1,\ldots,M$, where $d$ is the number of biomarkers. Suppose $\mathbf{X}_j \sim F_j$ for $j=1,\ldots, M$ where $F_j$ denotes a multivariate continuous distribution. Now the linear combination of the biomarkers corresponding to the $j$-th  outcome category is given by $\boldsymbol{\beta}^T\mathbf{X}_j = \sum_{k=1}^d\beta_kX_{jk},$ where $\boldsymbol{\beta} = (\beta_1,\ldots,\beta_d)$ denotes the combining coefficient vector. Without loss of generality, assuming that the higher value of the combination value corresponds to the higher outcome category, the HUM \citep{Li2008}, which measures the diagnostic accuracy, is given by
$D(\boldsymbol{\beta}) = P(\boldsymbol{\beta}^T\mathbf{X}_M > \cdots \boldsymbol{\beta}^T\mathbf{X}_2 > \boldsymbol{\beta}^T\mathbf{X}_1)$.
In order to distinguish the outcome categories, our goal is to find the optimal combination coefficient vector $\boldsymbol{\beta}_0$ for which $D(\boldsymbol{\beta})$ is maximized. So, $\boldsymbol{\beta}_0 = \argmax_{||\boldsymbol{\beta}|| = 1} D(\boldsymbol{\beta})$. Note that $D(\boldsymbol{\beta})$ should be maximized over all possible coefficient vectors of norm 1 to avoid the issue of non-identifiability. When $\mathbf{X}_1,\ldots, \mathbf{X}_M$ follow multivariate normal distributions, under a few regularity conditions, the value of $\boldsymbol{\beta}_0$ can be derived \citep{Su1993}. However, without any distributional assumption on $\mathbf{X}_1,\ldots, \mathbf{X}_M$, the value of $\boldsymbol{\beta}_0$ cannot be analytically obtained.

\subsection{Estimates of the HUM} 
We now review estimates of the HUM proposed 
in the literature. Essentially, these approaches provide different options for how to formulate the objective function.
\subsubsection*{Empirical hypervolume under the manifold (EHUM) } 
\
In order to estimate the optimal coefficient vector, \cite{Zhang2011} proposed maximizing the empirical estimate of the HUM from the given sample. Suppose the sample is denoted by
$\{\mathbf{X}_{ji_j}: j = 1,\ldots, M, i_j = 1,\ldots,n_j\}$, so
 the total sample size is $n = \sum_{j=1}^Mn_j$. Then the empirical estimate of HUM is given by 
\begin{equation*}
 {D}_{E}(\boldsymbol{\beta}) = \dfrac{1}{n_{1}n_{2} \cdots n_{M}} \displaystyle\sum_{i_{1}=1}^{n_{1}}\displaystyle\sum_{i_{2}=1}^{n_{2}} \cdots \displaystyle\sum_{i_{M}=1}^{n_{M}}I(\boldsymbol{\beta}^{T}\mathbf{X}_{i_{M}}>\boldsymbol{\beta}^{T}\mathbf{X}_{i_{(M-1)}}>\cdots>\boldsymbol{\beta}^{T}\mathbf{X}_{i_{1}}).
\end{equation*}
Here $I(\cdot)$ denotes the indicator function. The optimal combination coefficient vector obtained by maximizing ${D}_{E}(\boldsymbol{\beta})$ is given by $\hat{\boldsymbol{\beta}}_E = \argmax_{||\boldsymbol{\beta}|| = 1} D_E(\boldsymbol{\beta})$.
Note that $D_E(\boldsymbol{\beta})$ can be multi-modal. The discontinuity and non-differentiablity of $D_E(\boldsymbol{\beta})$ pose additional challenges in maximizing it using existing optimization methods.

\subsubsection*{Upper and lower bound approach (ULBA)}
\ 
In order to alleviate the computational burden of maximizing $D_E(\boldsymbol{\beta})$, \cite{Hsu2016} proposed alternative objective functions which can be maximized more easily. \cite{Hsu2016} showed that \\ \vskip-1cm
$$\max\{0, (M-1)P_{A}(\boldsymbol{\beta}) - (M-2)\} \le D(\boldsymbol{\beta}) \le P_{M}(\boldsymbol{\beta}),$$
where $P_{A}(\boldsymbol{\beta})$ and $P_{M}(\boldsymbol{\beta})$ are defined by
$$P_{A}(\boldsymbol{\beta}) =\frac{1}{M-1} \displaystyle\sum_{j=1}^{M-1}P(\boldsymbol{\beta}^{T}\mathbf{X}_{j+1}>\boldsymbol{\beta}^{T}\mathbf{X}_{j}), P_{M}(\boldsymbol{\beta}) = \min_{1 \le j \le M-1} P(\boldsymbol{\beta}^{T}\mathbf{X}_{j+1}>\boldsymbol{\beta}^{T}\mathbf{X}_{j}).$$
\cite{Hsu2016} proposed that instead of maximizing $D_E(\boldsymbol{\beta})$, we can either maximize $P_{A}(\boldsymbol{\beta})$ or $P_{M}(\boldsymbol{\beta})$, as they are much easier to solve.  Since they showed that the solution obtained by maximizing $P_A(\boldsymbol{\beta})$ yields better results than that obtained using $P_M(\boldsymbol{\beta})$, in this paper we consider the optimal coefficient combination vector using ULBA as given by $\hat{\boldsymbol{\beta}}_{ULBA} = \argmax_{||\boldsymbol{\beta}|| = 1} P_A(\boldsymbol{\beta})$. Despite the simpler form of $P_A(\boldsymbol{\beta})$ compared to $D_E(\boldsymbol{\beta})$, this function is still  discontinuous with possibly multiple modes.

\subsection{Existing techniques for estimating the optimal value of \texorpdfstring{$\boldsymbol{\beta}$}{beta}}
Due to the multi-modal nature of the objective functions discussed above, it is challenging to optimize the combination coefficient vector $\boldsymbol{\beta}$ simultaneously.
In Supplementary Material Section A, we provide  background on general-purpose global optimization techniques including the genetic algorithm (GA), simulated annealing (SA), and pattern search (PS) which can be applied here. In the current section, we focus on approaches designed specifically for estimation of the combination vector developed in the past few decades. 

 \cite{Pepe2006} proposed the step-down algorithm for combining multiple biomarkers. Although the method was first proposed for the binary categorical outcome case, this principle can also be used in scenarios with  more than two categorical outcomes \citep{Kang2013}. In the step-down approach, the biomarkers are first ordered based on their individual EHUM value. Then the coefficient of the biomarker with highest EHUM value is taken to be 1. Then at each step one additional biomarker is included and its coefficient is estimated. Thus, the coefficients of the biomarkers are estimated one at a time. A detailed description of the step-down algorithm is provided in Supplementary Material Section B.1. \cite{Hsu2016} used this algorithm to maximize the upper or lower bound of HUM, namely $P_{M}$ or $P_{A}$. 
 \cite{Liu2011} proposed the min-max (MM) principle in the context of a binary outcome, where, corresponding to each vector of biomarkers of length $d$, only the maximum and the minimum values of those $d$ values are used to estimate the combination coefficient vector. The estimation procedure using the MM principle is provided in detail in Supplementary Material Section B.2.

\section{Spherically Constrained Optimization Routine}
\label{sec_algo}
We now describe the algorithm for our proposed optimization approach, Spherically Constrained Optimization Routine (SCOR).
\subsection{Basic Principle}
\label{sec2_basic}

The basic principle of the search for the optimal value is as follows. Within an iteration, at first a step size $s>0$ is fixed. Then, two new points in the neighborhood are obtained by adding and subtracting  $s$ from each coordinate keeping all other coordinates fixed. Thus, at any given iteration, $2d$ possible new solution points are generated. For example, if $(\beta_1,\beta_2,\beta_3)$ denotes the current solution (taking $d=3$), then the new set of possible solutions  are $(\beta_1+s,\beta_2,\beta_3), (\beta_1-s,\beta_2,\beta_3), (\beta_1,\beta_2+s,\beta_3), (\beta_1,\beta_2-s,\beta_3), (\beta_1,\beta_2,\beta_3+s)$ and $(\beta_1,\beta_2,\beta_3-s)$. This strategy is related to the approach used in \cite{Fermi1952} to solve an unconstrained optimization problem.
This specific principle has two desirable properties: firstly, at each iteration the size of the search space (i.e., the set of newly generated points) is of $O(d)$ (i.e., $2d$), unlike GA, where the search space increases exponentially with the dimension \citep{Geris2012}. Secondly, after the step-size $s$ is fixed, the operations for finding the $2d$ new points and evaluating the objective function at those points are independent, so they can be performed in parallel. However, due to the spherical constraint on $\boldsymbol{\beta}$, this simple search strategy cannot be directly applied here.

\begin{figure}[]
	\centering
	\includegraphics[width=1\textwidth]{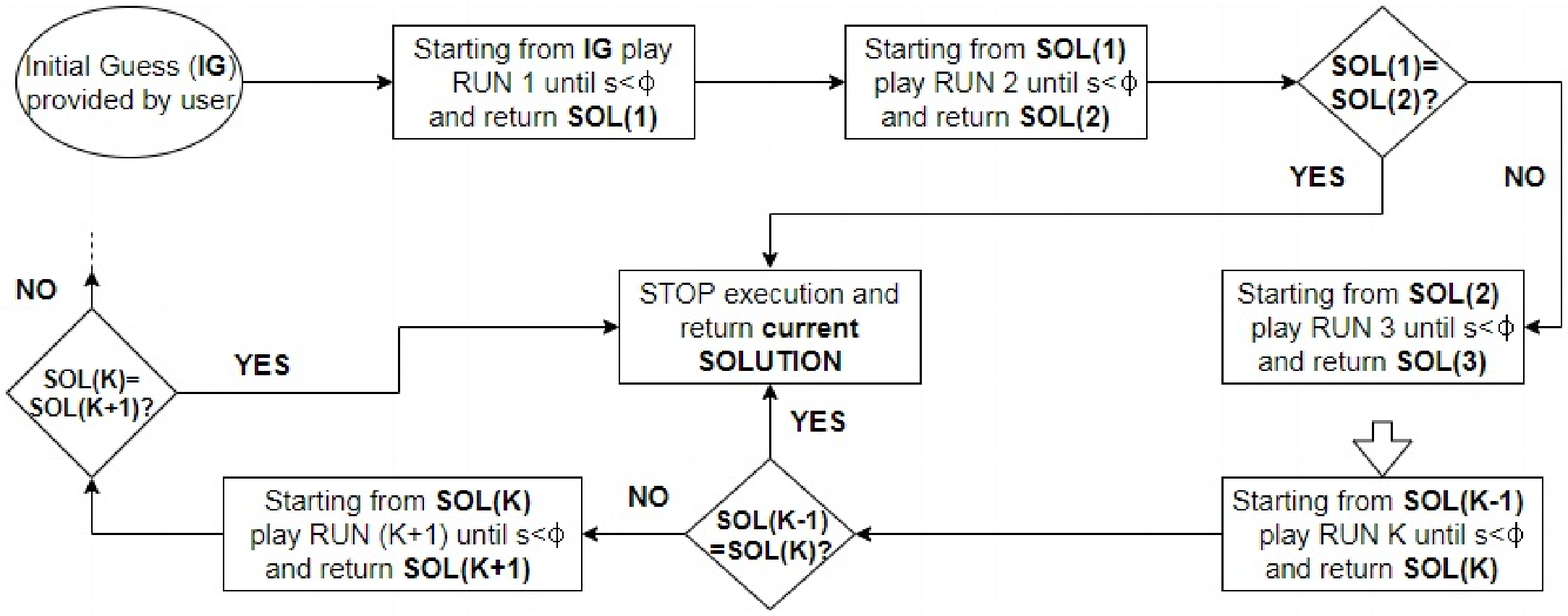}
	\caption{Flowchart of the SCOR algorithm where $s$ denotes the step size, $\phi$ denotes the step size threshold, and \textbf{SOL(k)} denotes the solution returned by the $k$-th \textit{run}.}
	\label{flowchart}
\end{figure}

We now summarize the proposed optimization scheme designed to address this issue.
The  algorithm can be broken into a sequence of \textit{runs} where within each \textit{run}, a number of iterations are performed. At the beginning of each iteration, the step-size $s>0$ is fixed. The location of the new set of points to be evaluated in the iteration depends on this step size $s$. The exact relationship between the step size and the new set of possible solutions is described in the following subsections.
The first iteration in the first \textit{run} starts from a starting point provided by the user. At each iteration, a new set of $2d$ points around the current solution is generated. As mentioned above, the location of these new points is directly related to the value of step-size $s$. In general, larger values of $s$ result in more distant points from the current solution. Then, the objective function is evaluated at all $2d+1$ points. At the end of an iteration, the point with the maximum objective function value is taken as the updated solution. Thus, at each iteration the objective function value either increases or stays the same. At the beginning of a \textit{run}, the step size is taken to be large. Depending on the improvement of the objective function across iterations, the step size is either reduced or kept same after each iteration. Once the step size within a \textit{run} becomes sufficiently small (determined by a threshold $\phi$ which is a tuning parameter of the algorithm), the \textit{run} ends, and the current solution is passed to to the next \textit{run}, which uses this solution as its starting point. Once two consecutive \textit{runs} yield the same solution, the algorithm stops execution and returns the final solution. A brief overview of the SCOR algorithm is given in Figure \ref{flowchart}.

\subsection{Derivation of adjustment step-size}
\label{sec_algo_derive}
 
\begin{figure}[]
	\centering
	\subfigure[]{\includegraphics[width=0.35\textwidth]{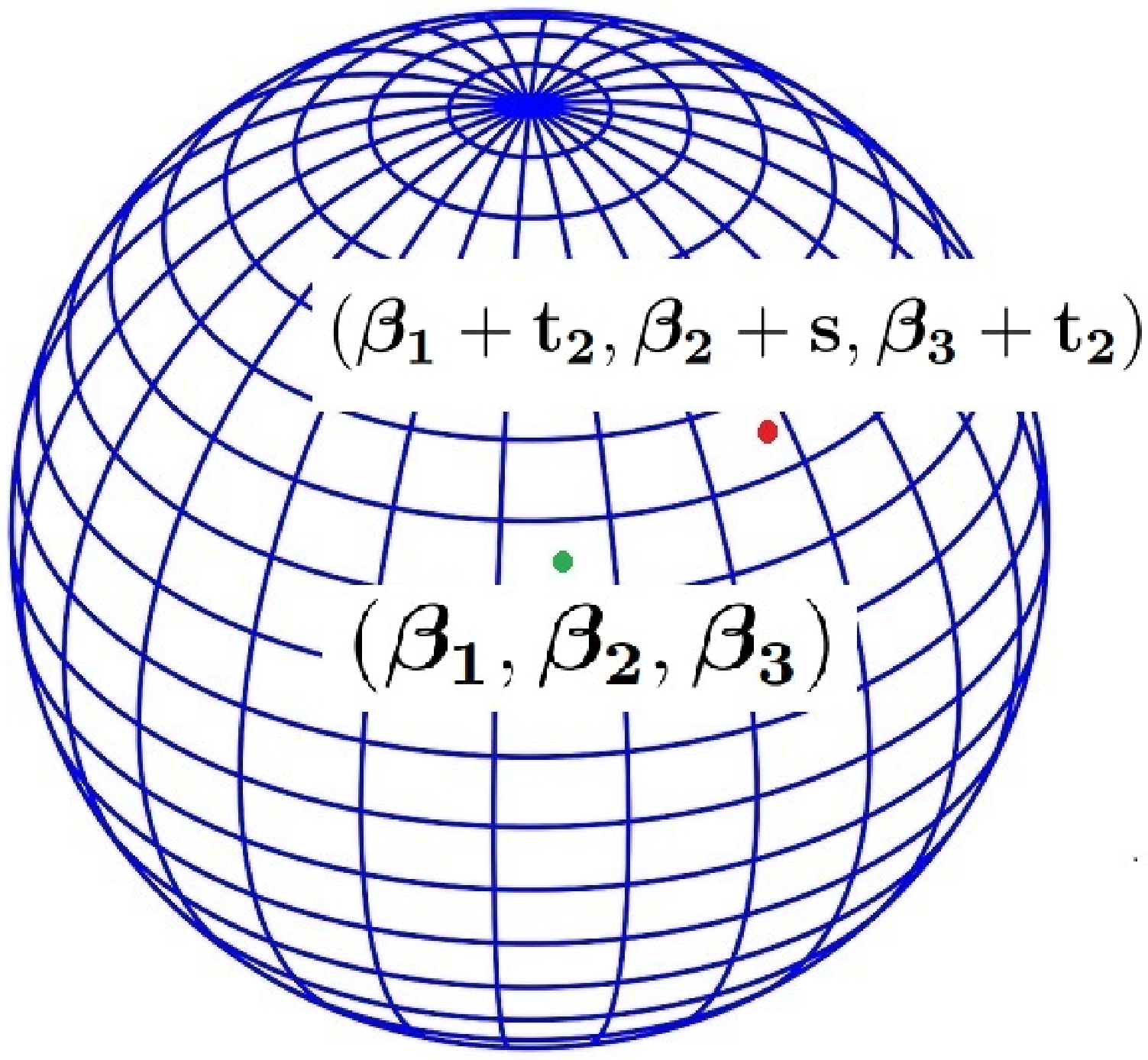}\label{fig:01} }
	\subfigure[]{\includegraphics[width=0.35\textwidth]{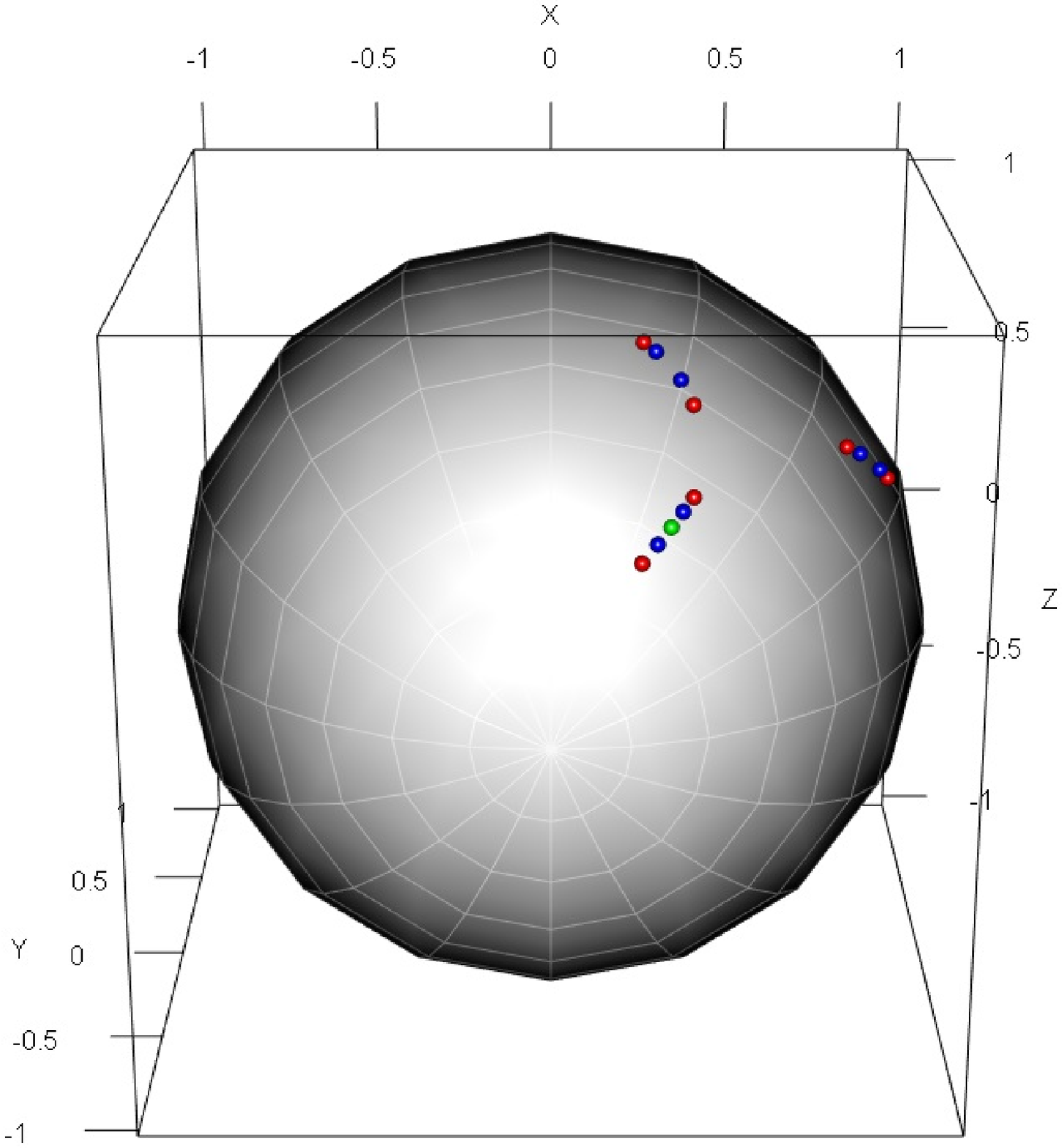}\label{fig:02} }
\caption{Exploratory moves on the surface of unit sphere by SCOR. Figure (a) shows an initial point $(\beta_1,\beta_2,\beta_3)$ (\textit{green}) on the surface of the unit sphere and a point explored  $(\beta_1+t_2,\beta_2+s,\beta_3+t_2)$ (\textit{red}) after making exploratory move with step-size $s$ for the $2^{nd}$ coordinate, where $t_2$ denotes the adjustment step-size. Figure (b) shows exploratory moves in a typical iteration of SCOR: starting from the current solution point $(0.289,-0.816,0.5)$ (\textit{green}), the points explored after making exploratory moves with step-size $s=0.03$ (\textit{blue}) and $s=0.06$ (\textit{red}).  }
\end{figure}

 The RMPS method \citep{Das2016} incorporates adjustments to the step size to ensure that the proposed points remain within a restricting hyper-rectangle.
In that approach, once the step size $s$ is added to (or subtracted from) any of the coordinates, the other coordinates are kept unchanged. However, over a spherically constrained parameter space, adding the step size $s$ to a coordinate of the current solution while keeping the other coordinates fixed would yield a point outside the unit sphere surface, assuming the current solution is on the unit sphere. To handle a spherically constrained parameter space, a critical challenge is devising an adjustment strategy such that when $s$ is added to (or subtracted from) the $i$-th coordinate, the adjustment of other coordinates ensures that the  proposed solution point is still on the unit sphere. We derive an appropriate adjustment step size $t_i$ which depends on $s$ and is chosen 
 so that when $t_i$ is added to the remaining coordinates, the  point obtained is still on the unit-sphere. In Figure \ref{fig:01} an exemplary plot is provided giving an idea why adjustment step-size should be a function of $s$.

For example, suppose within any given \textit{run}, at the $j$-th iteration, the current solution is given by $\boldsymbol{\beta}^{(j)} = (\beta_1^{(j)},\ldots,\beta_d^{(j)})$. Note that $\sum_{k=1}^d(\beta_k^{(j)})^2 = 1$. 
Suppose we update the $i$-th coordinate as $\beta_i^{(j+1)} = \beta_i^{(j)}+s.$ 
Then the corresponding adjustment step size $t_i$ is added to the rest of the coordinates, $\beta_k^{(j+1)} = \beta_k^{(j)}+t_i, \; k \in \{1,\ldots,d\} \setminus \{i\}.$ By the definition of the adjustment step size,
$\boldsymbol{\beta}^{(j+1)}$ is on the unit sphere as long as such a $t_i$ exists. Since $\boldsymbol{\beta}^{(j)}$ is also on the unit-sphere, we have,
\begin{align}
\sum_{k=1}^d (\beta_k^{(j)})^2 = \sum_{k=1}^d (\beta_k^{(j+1)})^2 =  \sum_{k=1,k\neq i}^d (\beta_k^{(j)} + t_i)^2 + (\beta_i^{(j)}+s)^2 = 1,
\end{align}
which implies,
\begin{align}
(d-1)t_i^2 + 2t_i\sum_{k=1,k\neq i}^d\beta_k^{(j)} + (2s\beta_i^{(j)} + s^2) =0.
\label{quad}
\end{align}
The value of $t_i$ can be obtained by solving \eqref{quad} to obtain $t_i = T_1(s)$ or $T_2(s)$ where 
\begin{align}
& T_1(s) = \frac{-2\sum_{k=1,k\neq i}^d\beta_k^{(j)} + \sqrt{D_i(s)}}{2(d-1)}, \quad
T_2(s) = \frac{-2\sum_{k=1,k\neq i}^d\beta_k^{(j)} - \sqrt{D_i(s)}}{2(d-1)}, \nonumber \\
& D_i(s) = \big(2\sum_{k=1,k\neq i}^d\beta_k^{(j)}\big)^2 - 4(d-1)(2s\beta_i^{(j)} + s^2).
\label{discriminant}
\end{align}
In order for the proposed algorithm to converge to the true solution, the value of $t_i$ needs to be chosen such that the distance of the new possible solution $\boldsymbol{\beta}^{(j+1)}$  from the current solution $\boldsymbol{\beta}^{(j)}$ should go to 0 as $s$ goes to 0, which only holds true for $t_i = T_1(s)$. So for a given step size $s$, the adjustment step size is taken to be $T_1(s)$. It should be noted that $T_1(s)$ may come out to be complex implying that for the given step size $s$ and coordinates of the current solution, no appropriate adjustment step size $t_i$ exists. In those cases, we adopt alternative strategies, as discussed in the Section \ref{sec_local}. In Figure \ref{fig:02}, starting with a current solution on the surface of unit sphere, the new set of explored points are plotted on the unit-sphere corresponding to two different exemplary step sizes. In order to make the algorithm more efficient, a minor modification is incorporated which is described in Section C of the Supplementary Material.
\subsection{Tuning parameters}
\label{sec_tuning}
In SCOR, the tuning parameters and their roles are similar to those considered in RMPS and RMPSS \citep{Das2016, Das2020}. 
Since the solution update strategy is consistent across \textit{runs}, explaining the roles of the tuning parameters within a single \textit{run} is sufficient to illustrate their effect. Within each \textit{run}, iterations are initialized with a starting point which is either the initial guess provided by the user (for the first \textit{run}) or the solution returned by the previous \textit{run} (for all \textit{run}s after the first \textit{run}). At the beginning, the step size is set to $s_{initial}$, which is preferably taken to be large in order to promote selection of distant candidate solutions w.r.t.\ the current solution. Suppose $\boldsymbol{\beta}^{(j)} = (\beta_1^{(j)},\;  \ldots, \beta_{d}^{(j)})$ denotes the solution at the end of $j$-th iteration. At the beginning of $(j+1)$-th iteration, suppose the current step-size is $s = s^{(j+1)} (>0)$. Then the set of $d$ new candidate solutions is given by $\{\boldsymbol{\beta}_{(i,+)}^{(j+1)}\}_{i=1}^d$ where 
\begin{align*}
\boldsymbol{\beta}^{(j+1)}_{(i,+)} = (\beta_1^{(j)}+t_i, \beta_2^{(j)}+t_i, \ldots, \beta_{i-1}^{(j)}+t_i, \beta_{i}^{(j)}+s,\beta_{i+1}^{(j)}+t_i, \ldots, \beta_{d}^{(j)}+t_i).
\end{align*}
for $s>0, i=1,\ldots,d$. Similarly, taking $s = -s^{(j+1)}$, another $d$ candidate solutions are generated given by $\{\boldsymbol{\beta}_{(i,-)}^{(j+1)}\}_{i=1}^d$ where
$$\boldsymbol{\beta}^{(j+1)}_{(i,-)} = (\beta_1^{(j)} + t_i^\prime , \beta_2^{(j)} + t_i^\prime, \ldots, \beta_{i-1}^{(j)}+t_i^\prime, \beta_{i}^{(j)}+s,\beta_{i+1}^{(j)}+t_i^\prime, \ldots, \beta_{d}^{(j)}+t_i^\prime), \; s<0.$$ Thus, within any given iteration, based on the step size $s$, $2d$ new candidate solutions are generated. The objective function is evaluated at these $2d+1$ points ($2d$ candidate solutions and the current solution $\boldsymbol{\beta}^{(j)}$), and the point with the lowest value of the objective function value is taken as the updated solution $\boldsymbol{\beta}^{(j+1)}$.
As discussed in Section \ref{sec_algo_derive}, in some scenarios, the adjustment step sizes $t_i$ or $t_i^\prime$ might be complex. In those settings, we propose alternative update strategies which is discussed in Section \ref{sec_local}.

Within a \textit{run}, at the end of any given iteration, if the improvement of the objective function value is less than a user-specified tolerance \textit{tol\_fun}, the step size is divided by a factor $\rho>1$ denoted as the \textit{step decay rate}.
So, at each iteration, the step size is either kept the same or reduced by dividing it by the \textit{step decay rate}.
For example, $s^{(j+1)}$ will be either $s^{(j)}$ or $\frac{s^{(j)}}{\rho}$ depending on the improvement to the solution obtained in the $(j+1)$-th iteration. The step size is reduced to enable finer search close to the current solution if no better solution is found in the set of candidate solutions using a larger value of $s$. The idea of  reducing the step size is a well-known strategy used in existing derivative-free optimization algorithms on unconstrained parameter spaces \citep{Fermi1952, Kerr2018}.

 We consider the step size as sufficiently close to 0 once its value gets smaller than the \textit{step size threshold} $\phi$. Once the step size becomes less than $\phi$, no further iterations are performed within that particular \textit{run}. If two consecutive \textit{run}s yield the same solution, the algorithm terminates after returning the solution obtained in the last \textit{run}.
 
 To handle cases where the solution is known to be sparse a priori, we consider the \textit{sparsity threshold} $\lambda$ as another tuning parameter. Once the step size for the $j$-th iteration $s^{(j)}$ is determined, before calculating the adjustment step size $t_i$, all the coefficients (except the $i$-th coordinate) with absolute values less than $\lambda$ are set equal to zero. Suppose at the $j$-th iteration, while updating $i$-th coordinate, out of the remaining $d-1$ coordinates, $h$ of them have absolute values less than $\lambda$. Then those $h$ coordinates are set equal to 0. The adjusted step size $t_i$ is calculated based on rest of the remaining $d-h$ coordinates. In case a sparse solution is not expected, the user can set the value of $\lambda$ to 0.

\subsection{Local step sizes}
\label{sec_local}
At the beginning of the $j$-th iteration, we initialize $2d$ \textit{local step sizes} such that $s_i^+=s^{(j)}$ and $s_i^-=-s^{(j)}$ for $i = 1,\ldots,d$. When adding $s_i^+$ (or $s_i^-$) to the $i$-th coordinate, if the corresponding adjustment step size $t_i$ (or $t_i^\prime$) exists, update movements are performed without modifying $s_i^+$ (or $s_i^-$). However, in case no such real $t_i$ (or $t_i^\prime$) exists for the given \textit{local step size} $s_i^+$ (or $s_i^-$), it is subsequently divided by the \textit{step decay rate} $\rho$ until the solution $t_i$ (or $t_i^\prime$)  becomes real. Note that in equation \eqref{discriminant}, $D_i(s) \xrightarrow{} \big(2\sum_{k=1,k\neq i}^d \beta_k^{(j)}\big)^2>0$ as $s \xrightarrow{} 0$. Therefore, a real $t_i$ (or $t_i^\prime$) will exist given a \textit{local step size} $s_i^+$ (or $s_i^-$) sufficiently close to 0 (see Theorem \ref{theorem1}). Since at the beginning of the iteration, the values of these \textit{local step-size}s are set equal to the step size for that iteration, the values of the \textit{local step size}s in the current iteration do not depend on the values from the previous iteration. 

\subsection{Overview of algorithm}
\label{sec_algo_steps}
Pseudocode for the proposed procedure is given in \textbf{Algorithm 1}. There we let $\hat{\boldsymbol{\beta}}^{(R)}$ denote the solution obtained at the end of $R$-th \textit{run}, and $\boldsymbol{\beta}^{(j)}$ denotes the solution obtained after the $j$-th iteration within a particular \textit{run}.
 We set an upper limit $max\_runs$ on the maximum number of \textit{run}s to be executed. Within each \textit{run}, there is an upper limit $max\_iters$ on the number of iterations. The roles of the tuning parameters along with their default values are provided in Supplementary Material Table S1.

\begin{algorithm}
\caption{SCOR}\label{euclid}
\begin{algorithmic}[1]
{\fontsize{8}{1}
\State $R \gets 1$
\vspace*{-5mm}
\BState \emph{top}:
\vspace*{-5mm}
\State $j \gets 1$
\vspace*{-5mm}
\State $s^{(0)} \gets s_{initial}$
\vspace*{-5mm}
\If {$R = 1$}
\vspace*{-5mm}
\State $\boldsymbol{\beta}^{(0)} \gets \text{Initial guess}$
\vspace*{-5mm}
\Else 
\vspace*{-5mm}
\State $\boldsymbol{\beta}^{(0)} \gets \hat{\boldsymbol{\beta}}^{(R-1)}$
\EndIf
\vspace*{-5mm}
\While {($j \leq max\_iter$ and $s^{(j)} > \phi$)}
\vspace*{-5mm}
\State $F \gets f(\boldsymbol{\beta}^{(j-1)})$
\vspace*{-5mm}
\State $s \gets s^{(j-1)}$
\vspace*{-5mm}
\For {$h = 1:2d$}
\vspace*{-5mm}
\State $i \gets [\frac{(h+1)}{2} ]$ ($[\cdot]$ denotes largest smaller integer function)
\vspace*{-5mm}
\State $\boldsymbol{\beta}_h \gets \boldsymbol{\beta}^{(j-1)}$
\vspace*{-5mm}
\State $\Lambda \gets \text{which}(|\beta_k^{(j-1)}| < \lambda), k \in \{1,\ldots,d\}
\setminus \{i\}$
\vspace*{-5mm}
\State $\Gamma \gets \text{which}(|\beta_k^{(j-1)}| \geq \lambda), k \in \{1,\ldots,d\}
\setminus \{i\}$
\vspace*{-5mm}
\State $ s_h \gets (-1)^hs^{(j)}$
\vspace*{-5mm}
\State $D \gets \big(2*sum(\boldsymbol{\beta}_h(\Gamma)))^2 - 4*length(\Gamma)*(2s_h\boldsymbol{\beta}_h(i) + s_h^2 - sumsquare(\boldsymbol{\beta}_h(\Lambda)))$.
\vspace*{-5mm}
\While {($D < 0$ and $|s_h|>\phi$)}
\vspace*{-5mm}
\State $s_h \gets \frac{s_h}{\rho}$
\vspace*{-5mm}
\State $D \gets \big(2*sum(\boldsymbol{\beta}_h(\Gamma)))^2 - 4*length(\Gamma)*(2s_h\boldsymbol{\beta}_h(i) + s_h^2 - sumsquare(\boldsymbol{\beta}_h(\Lambda)))$.
\vspace*{-5mm}
\EndWhile
\If {($D\geq 0$)}
\vspace*{-5mm}
\State $t \gets \frac{-2*sum(\boldsymbol{\beta}_h(\Gamma)) + \sqrt{D}}{2*length(\Gamma)}$
\vspace*{-5mm}
\State $\boldsymbol{\beta}_h(i) \gets \boldsymbol{\beta}_h(i) + s_h$
\vspace*{-5mm}
\State $\boldsymbol{\beta}_h(\Gamma) \gets \boldsymbol{\beta}_h(\Gamma) + t$
\vspace*{-5mm}
\State $\boldsymbol{\beta}_h(\Lambda) \gets 0$
\vspace*{-5mm}
\State $f_h \gets f(\boldsymbol{\beta}_h)$
\vspace*{-5mm}
\Else 
\vspace*{-5mm}
\State $f_h \gets F$
\vspace*{-5mm}
\EndIf
\EndFor
\State $\mathbf{B} \gets \argmin_{\boldsymbol{\beta}_h} f_h$
\vspace*{-5mm}
\State $FF \gets min(\{f_h\}_{h=1}^{2d})$ 
\vspace*{-5mm}
\If {($FF < F$)}
\vspace*{-5mm}
\State $\boldsymbol{\beta}^{(j)} \gets \mathbf{B}$
\vspace*{-5mm}
\Else
\vspace*{-5mm}
\State $\boldsymbol{\beta}^{(j)} \gets \boldsymbol{\beta}^{(j-1)}$
\vspace*{-5mm}
\EndIf
\If {($j > 1$)}
\vspace*{-5mm}
\If {($|F-min(F,FF)| < tol\_fun$ and $s>\phi$)}
\vspace*{-5mm}
\State $s \gets \frac{s}{\rho}$
\vspace*{-5mm}
\EndIf
\EndIf
\vspace*{-5mm}
\State $s^{(j)} \gets s$
\vspace*{-5mm}
\State $j \gets j+1$
\vspace*{-5mm}
\EndWhile
\State $\hat{\boldsymbol{\beta}}^{(R)} \gets \boldsymbol{\beta}^{(j)}$, 
\vspace*{-5mm}
\If {$||\hat{\boldsymbol{\beta}}^{(R)} - \hat{\boldsymbol{\beta}}^{(R-1)}|| < tol\_fun\_2$ }
\vspace*{-5mm}
\State \Return $\hat{\boldsymbol{\beta}} = \hat{\boldsymbol{\beta}}^{(R)}$ as \textbf{final solution}
\vspace*{-5mm}
\State \textbf{exit}
\vspace*{-5mm}
\Else
\vspace*{-5mm}
\State $R \gets R+1$
\vspace*{-5mm}
\State \textbf{goto} \emph{top}.
\item[]
\vspace*{-5mm}
\EndIf 
}
\end{algorithmic}
\end{algorithm}

\begin{theorem}
\label{theorem1}
 Suppose $\mathbf{S} = \{(x_1,\ldots,x_n) \in \mathrm{R}^n \; : \; \sum_{i=1}^n x_i^2 = 1, i=1,\cdots,n\}$. Consider a sequence of step-sizes $\delta_k = \frac{s}{\rho^k}$ for $k\in \mathrm{N}$ and $s\neq 0, \rho>1$. Then there exists a $K$ such that for $k \geq K$, all adjustment step sizes $\{t_i\}_{i=1}^d$ are real. 
  \end{theorem}

\begin{theorem}
\label{theorem2}
 Suppose $f$ is convex, continuous and differentiable on $\mathbf{S}$. Consider a sequence $\delta_k = \frac{s}{\rho^k}$ for $k\in \mathrm{N}$ and $s>0, \rho>1$. Suppose $\mathbf{u}$ is a point in $\mathbf{S}$. Define $\mathbf{u}_k^{(i+)} = (u_1+t_i(\delta_k),\ldots, u_{i-1}+t_i(\delta_k), u_i+\delta_k,u_{i+1}+t_i(\delta_k), \ldots, u_n+t_i(\delta_k))$ and $\mathbf{u}_k^{(i-)} = (u_1+t_i(-\delta_k),\ldots, u_{i-1}+t_i(-\delta_k),u_i-\delta_k,u_{i+1}+t_i(-\delta_k), \ldots, u_n+t_i(-\delta_k))$ for $i=1,\cdots,n$, where $t_i(s)$ denotes the adjustment step size corresponding to step size $s$.
Given conditions detailed in Supplementary Material Section D, the global minimum of $f$ over $\mathbf{S}$ occurs at $\mathbf{u}$. 
 \end{theorem}
 
\noindent The proofs of Theorems \ref{theorem1} and \ref{theorem2} are provided in Supplementary Material Section D. As is the case for any existing black-box optimization method, SCOR is not theoretically guaranteed to find the global minimum of any arbitrary black-box function. However, it incorporates strategies to avoid getting stuck at local minima. As shown in Section E of the Supplementary Material, SCOR outperform competing black-box optimization methods including GA, SA, and PS in terms of identifying better solutions for benchmark functions. Moreover, as discussed in the next section, in the context of maximizing HUM estimates, SCOR outperforms existing methods proposed for biomarker combination as well.
\section{Simulation study}
\label{sec_sim}
In this section, we compare the performance of SCOR with that of the Nelder-Mead (NM), step-down, and min-max methods in the context of simulated data.
Under different simulation scenarios, we estimate the optimal combination vector by maximizing the the EHUM and ULBA objective functions using each of these four approaches.  We measure performance in terms of the EHUM objective function value $D_E(\cdot)$ at the estimated optimal solution, evaluated on a separate test data set. The SCOR algorithm is implemented in MATLAB 2016b. NM is an unconstrained optimization technique, so in applying this algorithm, we set the first parameter value to be either 1 or -1, whichever maximizes the individual EHUM if we set other parameter values to be 0. Then, the rest of the parameters are estimated by maximizing EHUM and ULBA using NM, performed using the \texttt{fminsearch} function in MATLAB 2016b. For the step-down and min-max algorithms also, we use the \texttt{fminsearch} function with default options.

We consider three simulation scenarios, two based on the normal distribution, and one based on the Weibull distribution (to illustrate performance in the context of non-normal data). Among the two  normal distribution scenarios considered, in the first one, we simulate covariates that  are not correlated, while in the second, the covariates have non-zero correlation. For each simulation scenario, we consider settings with $M=2$ and $M=3$ ordinal outcomes. The number of biomarkers is taken to be $d = 5,10,$ or $20$. In each case, the $d$ biomarkers are generated from the corresponding normal or Weibull distribution. For the case of $M=2$ categories, we set the sample sizes for each category to be 15, 30, and 60. For the $M=3$ category case, we take the sample sizes of each category to be 15 and 30. \\ \vskip-.3cm
\noindent \textbf{Scenario 1:} For the $i$-th disease category, the values of the biomarkers  are simulated from the $d$-variate normal distribution for $d=5,10,$ or $15$ with mean $\boldsymbol{\mu}_{i} = \big(\mu_{i1},\ldots,\mu_{id}\big)$, and covariance matrix $\mathbf{\Sigma}_i=\mathbf{I}_d$, where $\mu_{ij}=(-1)^j*i[1+0.1(j-1)]$, for $j = 1,\ldots,d$, $i=0,1$ (for the two category outcome case) and $i=0,1,2$ (for the three category outcome case). \\ \vskip-.5cm
\noindent \textbf{Scenario 2:} In order to explore the relative performance of all methods in cases with correlated predictors, here we consider the same model as Scenario 1 except we take $\mathbf{\Sigma}_i=(a_{st})_{d\times d}$ where $a_{st} = (0.5)^{|s-t|}$ for $s,t = 1,\ldots,d$ and $i=0,1$ (for the two category outcome case) and $i=0,1,2$ (for the three category outcome case). \\ \vskip-.3cm
\noindent \textbf{Scenario 3:} Here the values of the biomarkers are generated from the multivariate Weibull distribution. For the $i$-th disease category, the $j$-th biomarker follows a univariate Weibull distribution with scale parameter $\lambda_i$, shape parameter $k_j$ and location parameter $\gamma_j$.
We take $\lambda_i = i+1,k_j=0.5*j,\gamma_j = (-5)^j$ for $j = 1,\ldots,d$, $i=0,1$ (for the two category outcome case) and $i=0,1,2$ (for the three category outcome case).\\\\
\noindent
 For any given  scenario using any of the methods considered, we first estimate the biomarker coefficient vector maximizing EHUM and ULBA. We then generate a new sample of the same size under the same simulation design. The EHUM values reported (i.e., the objective function value $D_E(\cdot)$) are obtained using the biomarker coefficient vectors estimated on the training data on the new test set.
We adopt this strategy to ensure that the performance estimates do not reflect overfitting to the training data. We report the mean EHUM at the estimated optimal solutions using the SCOR, NM, step-down and min-max algorithms for the different simulation scenarios over 100 replications.

\begin{table}[h!]
\resizebox{1\columnwidth}{!}{%
\begin{tabular}{ccccccccc}
\hline
\multirow{2}{*}{\begin{tabular}[c]{@{}c@{}}Scenarios \end{tabular}} & \multirow{2}{*}{$d$} & \multirow{2}{*}{Method} & \multicolumn{2}{c}{Scenario 1} & \multicolumn{2}{c}{Scenario 2} & \multicolumn{2}{c}{Scenario 3} \\ \cline{4-9} 
 &  &  & ULBA & EHUM & ULBA & EHUM & ULBA & EHUM \\ \hline
\multirow{12}{*}{\begin{tabular}[c]{@{}c@{}}$M=2,$\\  $N = (15,15)$\end{tabular}} & \multirow{4}{*}{5} & SCOR & \textbf{0.928 (0.06)} & \textbf{0.928 (0.06)} & \textbf{0.974 (0.04)} & \textbf{0.974 (0.04)} & \textbf{0.900 (0.08)} & \textbf{0.900 (0.08)} \\
 &  & NM & 0.806 (0.16) & 0.806 (0.16) & 0.889 (0.14) & 0.889 (0.14) & 0.760 (0.11) & 0.760 (0.11) \\
 &  & Step-Down & 0.538 (0.30) & 0.538 (0.30) & 0.536 (0.31) & 0.536 (0.31) & 0.866 (0.08) & 0.866 (0.08) \\
 &  & Min-Max & 0.691 (0.10) & 0.691 (0.10) & 0.711 (0.13) & 0.711 (0.13) & 0.839 (0.08) & 0.839 (0.08) \\ \cline{2-9} 
 & \multirow{4}{*}{10} & SCOR & \textbf{0.972 (0.04)} & \textbf{0.972 (0.04)} & \textbf{0.984 (0.03)} & \textbf{0.984 (0.03)} & 0.953 (0.06) & 0.953 (0.06) \\
 &  & NM & 0.864 (0.15) & 0.864 (0.15) & 0.834 (0.18) & 0.834 (0.18) & 0.866 (0.09) & 0.866 (0.09) \\
 &  & Step-Down & 0.503 (0.28) & 0.503 (0.28) & 0.581 (0.32) & 0.581 (0.32) & 0.971 (0.05) & 0.971 (0.05) \\
 &  & Min-Max & 0.881 (0.07) & 0.881 (0.07) & 0.860 (0.08) & 0.860 (0.08) & \textbf{0.982 (0.02)} & \textbf{0.982 (0.02)} \\ \cline{2-9} 
 & \multirow{4}{*}{20} & SCOR & \textbf{0.971 (0.04)} & \textbf{0.971 (0.04)} & \textbf{0.979 (0.04)} & \textbf{0.979 (0.04)} & 0.958 (0.06) & 0.958 (0.06) \\
 &  & NM & 0.805 (0.25) & 0.805 (0.25) & 0.692 (0.26) & 0.692 (0.26) & 0.930 (0.07 & 0.930 (0.07 \\
 &  & Step-Down & 0.472 (0.33) & 0.472 (0.33) & 0.472 (0.32) & 0.472 (0.32) & 0.975 (0.04) & 0.975 (0.04) \\
 &  & Min-Max & 0.922 (0.05) & 0.922 (0.05) & 0.909 (0.06) & 0.909 (0.06) & \textbf{0.997 (0.01)} & \textbf{0.997 (0.01)} \\ \hline
\multirow{12}{*}{\begin{tabular}[c]{@{}c@{}}$M=2,$\\  $N = (30,30)$\end{tabular}} & \multirow{4}{*}{5} & SCOR & \textbf{0.950 (0.04)} & \textbf{0.950 (0.04)} & \textbf{0.982 (0.03)} & \textbf{0.982 (0.03)} & \textbf{0.923 (0.04)} & \textbf{0.923 (0.04)} \\
 &  & NM & 0.931 (0.03) & 0.931 (0.03) & 0.970 (0.03) & 0.970 (0.03) & 0.773 (0.10) & 0.773 (0.10) \\
 &  & Step-Down & 0.596 (0.28) & 0.596 (0.28) & 0.590 (0.30) & 0.590 (0.30) & 0.893 (0.06) & 0.893 (0.06) \\
 &  & Min-Max & 0.804 (0.10) & 0.804 (0.10) & 0.873 (0.12) & 0.873 (0.12) & 0.865 (0.05) & 0.865 (0.05) \\ \cline{2-9} 
 & \multirow{4}{*}{10} & SCOR & \textbf{0.990 (0.01)} & \textbf{0.990 (0.01)} & \textbf{0.995 (0.01)} & \textbf{0.995 (0.01)} & \textbf{0.989 (0.02)} & \textbf{0.989 (0.02)} \\
 &  & NM & 0.978 (0.02) & 0.978 (0.02) & 0.983 (0.03) & 0.983 (0.03) & 0.872 (0.08) & 0.872 (0.08) \\
 &  & Step-Down & 0.546 (0.32) & 0.546 (0.32) & 0.473 (0.32) & 0.473 (0.32) & 0.982 (0.03) & 0.982 (0.03) \\
 &  & Min-Max & 0.926 (0.05) & 0.926 (0.05) & 0.945 (0.06) & 0.945 (0.06) & 0.982 (0.02) & 0.982 (0.02) \\ \cline{2-9} 
 & \multirow{4}{*}{20} & SCOR & \textbf{0.991 (0.01)} & \textbf{0.991 (0.01)} & \textbf{0.984 (0.10)} & \textbf{0.984 (0.10)} & 0.985 (0.02) & 0.985 (0.02) \\
 &  & NM & 0.991 (0.01) & 0.991 (0.01) & 0.992 (0.01) & 0.992 (0.01) & 0.929 (0.05) & 0.929 (0.05) \\
 &  & Step-Down & 0.546 (0.32) & 0.546 (0.32) & 0.515 (0.31) & 0.515 (0.31) & 0.984 (0.04) & 0.984 (0.04) \\
 &  & Min-Max & 0.961 (0.04) & 0.961 (0.04) & 0.972 (0.04) & 0.972 (0.04) & \textbf{0.997 (0.01)} & \textbf{0.997 (0.01)} \\ \hline
\multirow{12}{*}{\begin{tabular}[c]{@{}c@{}}$M=3,$\\  $N = (15,15,15)$\end{tabular}} & \multirow{4}{*}{5} & \multicolumn{1}{l}{SCOR} & \multicolumn{1}{l}{\textbf{0.891 (0.10)}} & \multicolumn{1}{l}{\textbf{0.890 (0.10)}} & \multicolumn{1}{l}{\textbf{0.955 (0.08)}} & \multicolumn{1}{l}{\textbf{0.955 (0.08)}} & \multicolumn{1}{l}{\textbf{0.722 (0.01)}} & \multicolumn{1}{l}{\textbf{0.719 (0.10)}} \\
 &  & \multicolumn{1}{l}{NM} & \multicolumn{1}{l}{0.829 (0.11)} & \multicolumn{1}{l}{0.837 (0.11)} & \multicolumn{1}{l}{0.898 (0.13)} & \multicolumn{1}{l}{0.902 (0.13)} & \multicolumn{1}{l}{0.484 (0.12)} & \multicolumn{1}{l}{0.485 (0.12)} \\
 &  & \multicolumn{1}{l}{Step-Down} & \multicolumn{1}{l}{0.358 (0.32)} & \multicolumn{1}{l}{0.356 (0.32)} & \multicolumn{1}{l}{0.398 (0.38)} & \multicolumn{1}{l}{0.398 (0.38)} & \multicolumn{1}{l}{0.638 (0.12)} & \multicolumn{1}{l}{0.644 (0.12)} \\
 &  & \multicolumn{1}{l}{Min-Max} & \multicolumn{1}{l}{0.582 (0.15)} & \multicolumn{1}{l}{0.582 (0.15)} & \multicolumn{1}{l}{0.703 (0.23)} & \multicolumn{1}{l}{0.703 (0.23)} & \multicolumn{1}{l}{0.582 (0.10)} & \multicolumn{1}{l}{0.582 (0.10)} \\ \cline{2-9} 
 & \multirow{4}{*}{10} & \multicolumn{1}{l}{SCOR} & \multicolumn{1}{l}{\textbf{0.978 (0.03)}} & \multicolumn{1}{l}{\textbf{0.967 (0.10)}} & \multicolumn{1}{l}{\textbf{0.985 (0.03)}} & \multicolumn{1}{l}{\textbf{0.986 (0.03)}} & \multicolumn{1}{l}{\textbf{0.938 (0.06)}} & \multicolumn{1}{l}{\textbf{0.942 (0.05)}} \\
 &  & \multicolumn{1}{l}{NM} & \multicolumn{1}{l}{0.916 (0.11)} & \multicolumn{1}{l}{0.907 (0.14)} & \multicolumn{1}{l}{0.912 (0.19)} & \multicolumn{1}{l}{0.925 (0.19)} & \multicolumn{1}{l}{0.664 (0.13)} & \multicolumn{1}{l}{0.664 (0.13)} \\
 &  & \multicolumn{1}{l}{Step-Down} & \multicolumn{1}{l}{0.354 (0.34)} & \multicolumn{1}{l}{0.349 (0.34)} & \multicolumn{1}{l}{0.329 (0.36)} & \multicolumn{1}{l}{0.329 (0.36)} & \multicolumn{1}{l}{0.902 (0.08)} & \multicolumn{1}{l}{0.903 (0.08)} \\
 &  & \multicolumn{1}{l}{Min-Max} & \multicolumn{1}{l}{0.825 (0.09)} & \multicolumn{1}{l}{0.824 (0.09)} & \multicolumn{1}{l}{0.823 (0.13)} & \multicolumn{1}{l}{0.823 (0.13)} & \multicolumn{1}{l}{0.887 (0.05)} & \multicolumn{1}{l}{0.887 (0.05)} \\ \cline{2-9} 
 & \multirow{4}{*}{15} & \multicolumn{1}{l}{SCOR} & \multicolumn{1}{l}{\textbf{0.977 (0.03)}} & \multicolumn{1}{l}{\textbf{0.927 (0.22)}} & \multicolumn{1}{l}{\textbf{0.985 (0.02)}} & \multicolumn{1}{l}{\textbf{0.945 (0.20)}} & \multicolumn{1}{l}{0.952 (0.06)} & \multicolumn{1}{l}{0.951 (0.06)} \\
 &  & \multicolumn{1}{l}{NM} & \multicolumn{1}{l}{0.935 (0.18)} & \multicolumn{1}{l}{0.888 (0.27)} & \multicolumn{1}{l}{0.914 (0.20)} & \multicolumn{1}{l}{0.853 (0.29)} & \multicolumn{1}{l}{0.795 (0.09)} & \multicolumn{1}{l}{0.795 (0.09)} \\
 &  & \multicolumn{1}{l}{Step-Down} & \multicolumn{1}{l}{0.297 (0.35)} & \multicolumn{1}{l}{0.308 (0.35)} & \multicolumn{1}{l}{0.322 (0.35)} & \multicolumn{1}{l}{0.322 (0.35)} & \multicolumn{1}{l}{0.935 (0.08)} & \multicolumn{1}{l}{0.935 (0.08)} \\
 &  & \multicolumn{1}{l}{Min-Max} & \multicolumn{1}{l}{0.895 (0.06)} & \multicolumn{1}{l}{0.895 (0.06)} & \multicolumn{1}{l}{0.889 (0.08)} & \multicolumn{1}{l}{0.889 (0.08)} & \multicolumn{1}{l}{\textbf{0.964 (0.03)}} & \multicolumn{1}{l}{\textbf{0.964 (0.03)}} \\ \hline
\end{tabular}}
\caption{Performance comparison for two and three ordinal outcomes, where each class has sample size 15 and 30 for the two category case, and 15 for the three category case. The empirical hypervolume under manifolds (EHUM) and upper and lower bound approach (ULBA) objective functions are maximized by the proposed Spherically Constrained Optimization Routine (SCOR) algorithm and the existing Nelder-Mead (NM), step-down, and min-max algorithms. The estimated biomarker coefficient vectors are then used to calculate the EHUM value on a new dataset of the same size generated from the corresponding model. The entire procedure is repeated 100 times, resulting in 100 simulated training and test data sets, and the mean EHUM objective function values on the test data are reported, with the standard error in the parentheses. The result for the method with the best performance is marked in bold.} 
\label{category_all}
\end{table}
 In Table \ref{category_all}, we provide the results for the two category outcome setting for sample sizes 15 and 30 per group and for the three category outcome setting with sample size 15 per group. The results for the two category outcome setting with sample size 60 and for the three category outcome setting with sample size 30 are provided in Table S3 of the Supplementary Material.  SCOR outperforms both NM and step-down for all the scenarios considered. SCOR also  outperforms the min-max method for all scenarios except for a few cases for Scenario 3 with higher sample sizes; however, in those cases, the SCOR performance is quite close to that obtained using min-max. As the dimension $d$ of the biomarker vector increases, estimates by SCOR tend to improve, while the step-down algorithm tends to underperform. As the dimension $d$ increases, SCOR tends to outperform step-down by a higher margin. In general, the SCOR estimates  have lower standard errors compared to other methods, indicating more stable estimates, which is expected when using a global optimization approach.
 
In Table S2 of the Supplementary Material, we compare the performance of SCOR with general-purpose global optimization algorithms based on optimization of five benchmark functions on the unit-spherical parameter space with dimensions $d=5,20,50,100,$ and $500$, and show that in general SCOR outperforms existing optimization algorithms, with greatly reduced computation time. Specifically, 
using SCOR,  we obtain an improvement in in computation time up to 67 fold over pattern search, up to 43 fold over simulated annealing, and up to 38 fold over the genetic algorithm.

\section{Prediction of swallowing difficulty following radiation therapy}

\label{sec_appl}
In this section, we illustrate the SCOR method with an application to data from a study of swallowing difficulty, or dysphagia, in patients treated with radiation therapy for oropharyngeal cancer.
Outcomes of observational studies and clinical trials in cancer are commonly defined as categorical or ordinal variables, and
standardized systems have been adopted for both response and toxicity assessment in this framework.
 Specifically, the standard definition of response to cancer therapy in solid tumors  includes four categories reflecting the treatment efficacy: complete response, partial response, stable disease, and progressive disease \citep{Therasse2000}. Toxicities experienced as a consequence of treatment also follow a standard definition for classification and grading, the Common Terminology for Criteria for Adverse Events \citep[CTCAE,][]{Trotti2003}. The categories on this scale are 0 for no symptoms, 1 for mild, 2 for moderate, 3 for severe, 4 for life-threatening, and 5 for death. Similar scales have  been developed for outcomes of interest in specific cancer types: in particular, the DIGEST scale  measures swallowing difficulty using a similar grading scale to that of CTCAE \citep{Hutcheson2017, Goepfert2018}.
 
 In this case study, we analyze data derived from a study of swallowing difficulties, quantified using the DIGEST scale, as a result of radiation therapy for oropharyngeal cancer \citep{Kamal2018}. Radiation therapy is commonly used to treat oropharyngeal cancer \citep{Daly2010}. Although this treatment is critical to ensure control of the tumor, radiation to structures in the head and neck related to swallowing can lead to loss of function, and ultimately
 reduced loss of quality of life or serious complications, such as aspiration-related pneumonia \citep{Eisbruch2002, Hutcheson2014}. Understanding the impact of radiation dose to specific structures on swallowing outcomes can inform the development of radiation treatment plans designed to minimize dose to critical structures \citep{Duprez2013}.

In our dataset, radiation dose is characterized as the mean dose delivered to 13 critical structures. Here we focus on the 87 subjects with no missing values for dose. The response variable describing swallowing difficulty was quantified using the DIGEST scale; to ensure sufficient sample size in each class, we collapsed grades 2-4 into a single group \citep{Hutcheson2017}. The sample sizes per class are then 29 subjects with no dysphagia (0 on the DIGEST scale), 27 with mild dysphagia (1 on the DIGEST scale), and 30 with moderate to severe dysphagia (2-4 on the DIGEST scale).
Since some of the predictors were highly correlated, we performed a screening step to filter out the most highly correlated using a greedy approach, with variables that had the highest average correlation removed from the set until all pairwise correlations were less than 0.8. After screening out 4 variables, 
we had 9 remaining predictors: mean radiotherapy dose to cricopharyngeal muscle, esophagus, glottic area, inferior pharyngeal constrictor (PC), ipsilateral anterior digastric muscle (ADM), middle PC, mylohoid muscle, superior PC, and intrinsic tongue muscles.


To illustrate the relative performance of the SCOR, step-down and min-max techniques, we compute the estimated combination coefficients maximizing the ULBA and EHUM criteria using each method. The EHUM value at those solutions are then evaluated. We also compute the optimal cut-points to categorize the combination scores (obtained by multiplying the optimal coefficient vector with the biomarker values) using Youden's index \citep{Youden1950}. Youden's index is defined as the maximum possible value of $(\text{sensitivity}+ \text{specificity} - 1)$ over all possible decision thresholds, and provides a summary measure combining sensitivity and specificity \citep{Luo2012}. Note that before obtaining the combination scores, all the solution combination vectors obtained in the step-down and min-max techniques are divided by their corresponding norms so that each solution vector has norm 1.

\begin{figure}[]
	\centering
	\subfigure[ULBA (SCOR) \newline $D_E=\mathbf{0.419}, YI = \mathbf{0.471}$] {\includegraphics[width=0.23\textwidth]{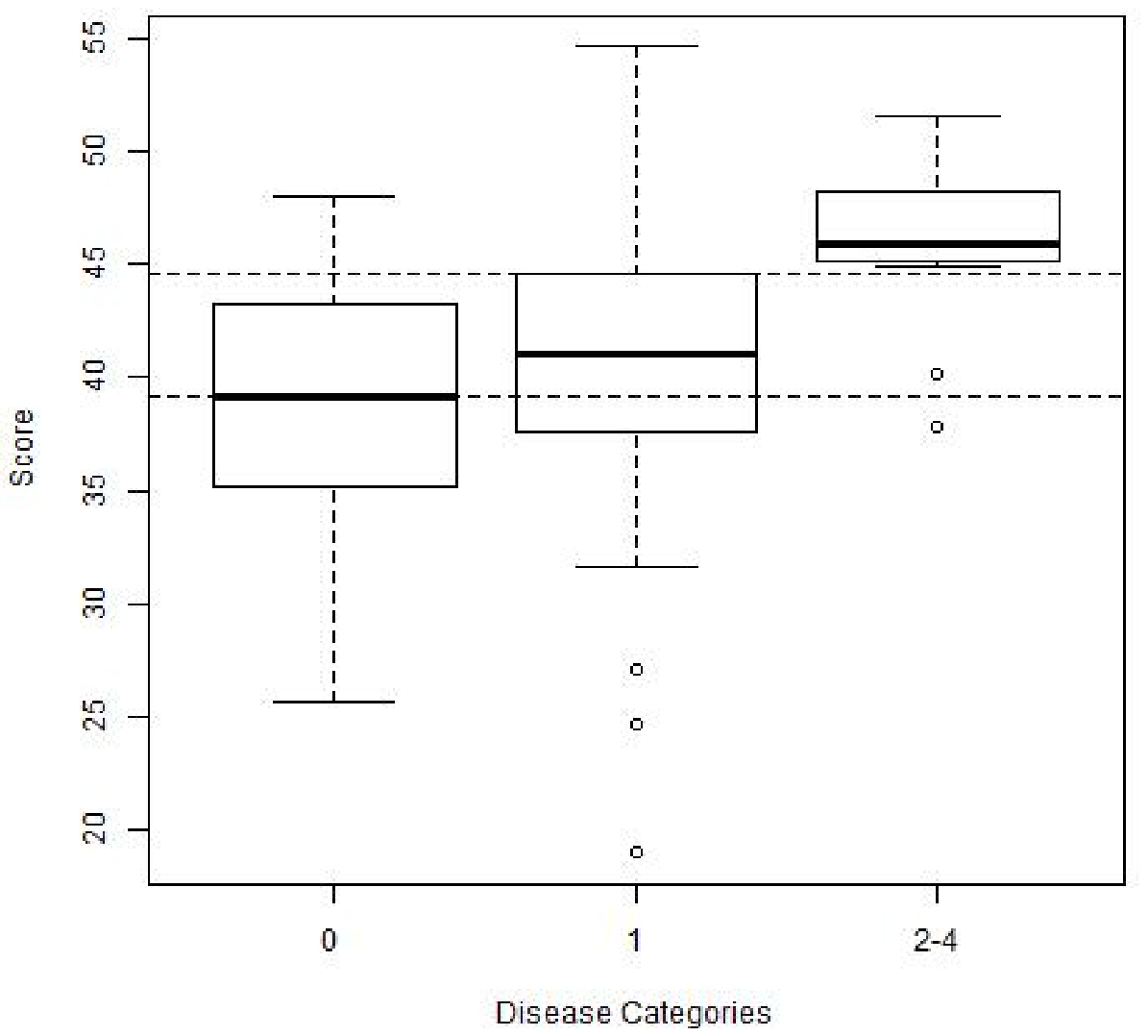}\label{fig:box01} }
	 \subfigure[ULBA (NM) \newline $D_E=0.341,YI=0.334$] {\includegraphics[width=0.23\textwidth]{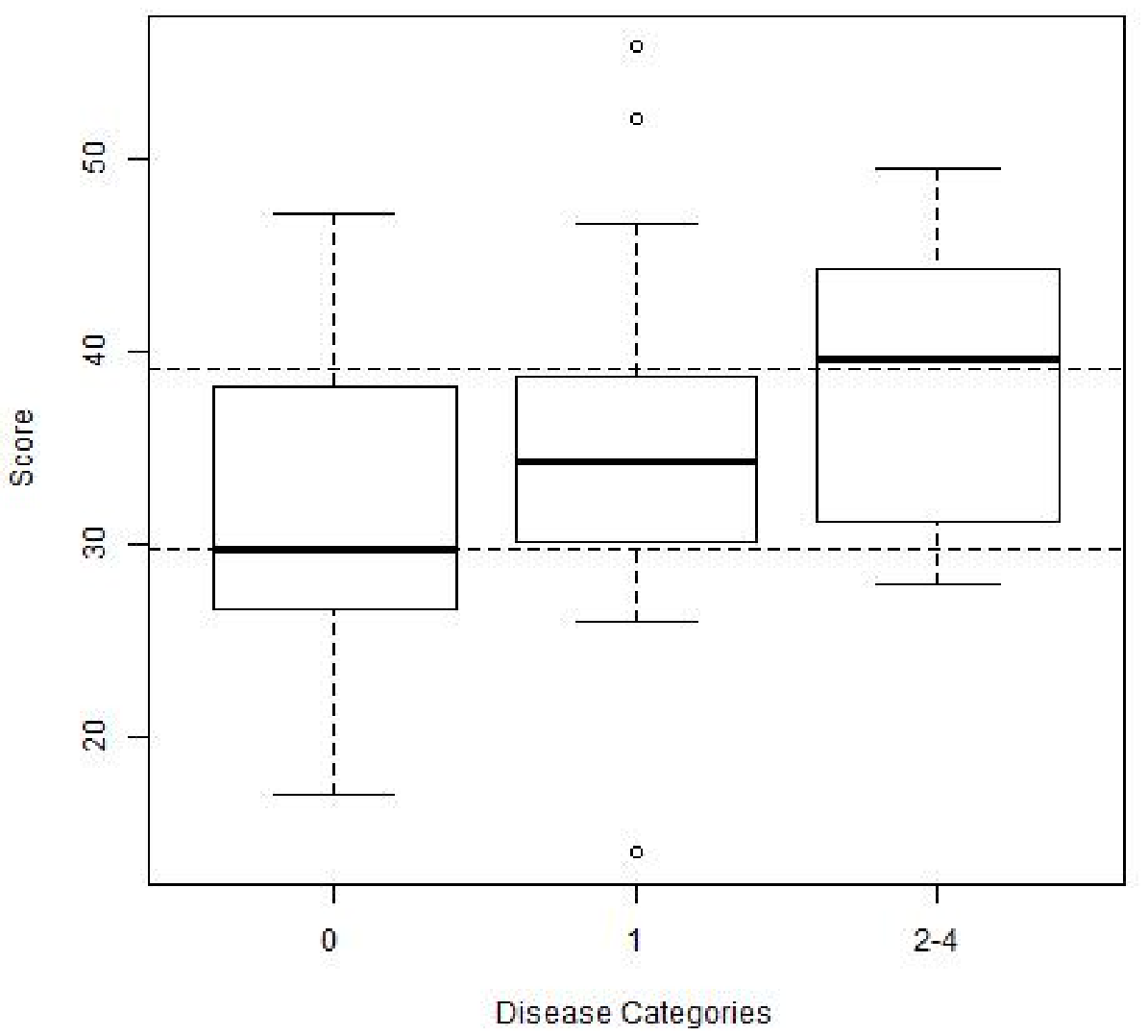}\label{fig:box02} }
	 \subfigure[ULBA (Step-down) \newline $D_E = 0.399, YI = 0.408$] {\includegraphics[width=0.23\textwidth]{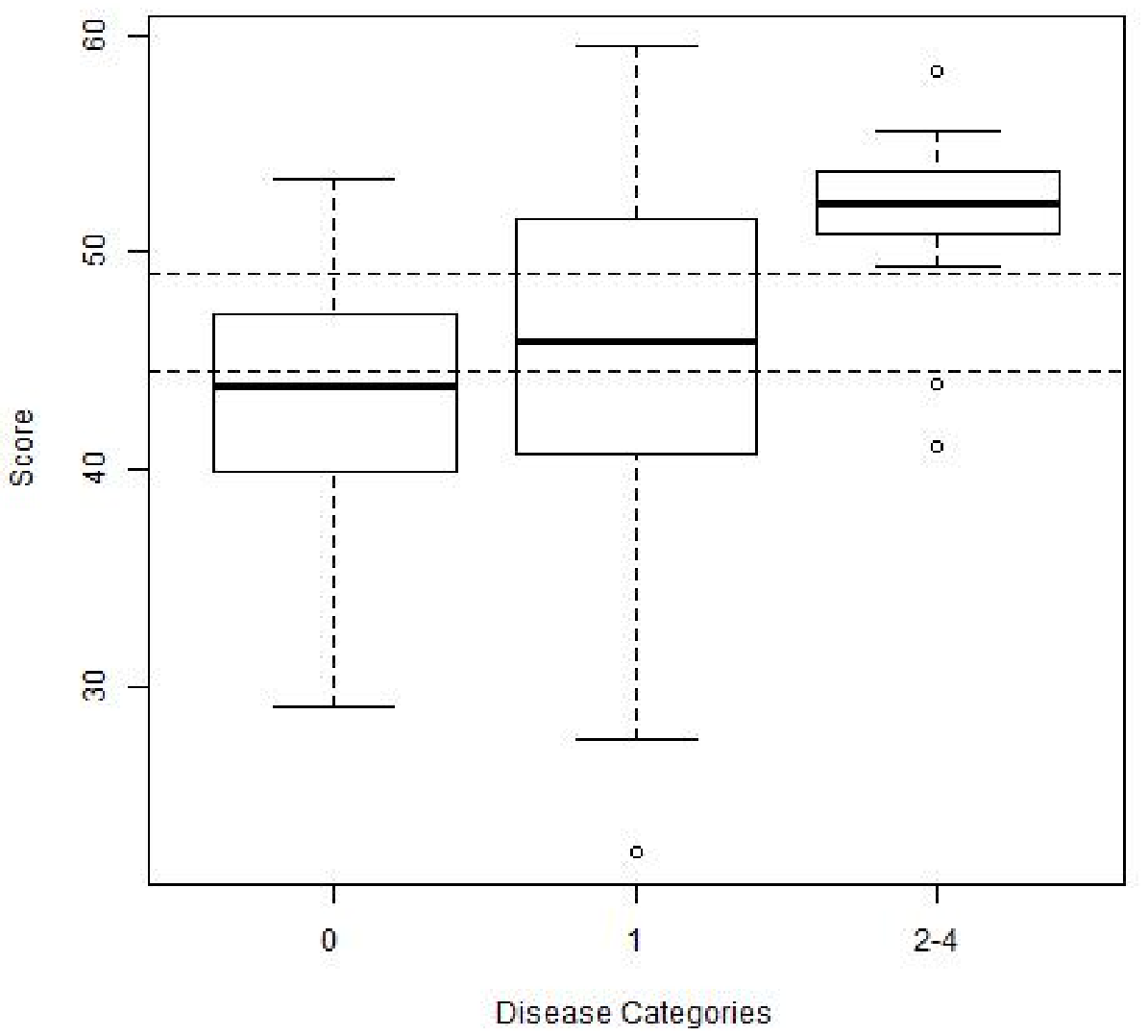}\label{fig:box03} }
	 \subfigure[ULBA (Min-max) \newline $D_E = 0.363, YI = 0.361$] {\includegraphics[width=0.23\textwidth]{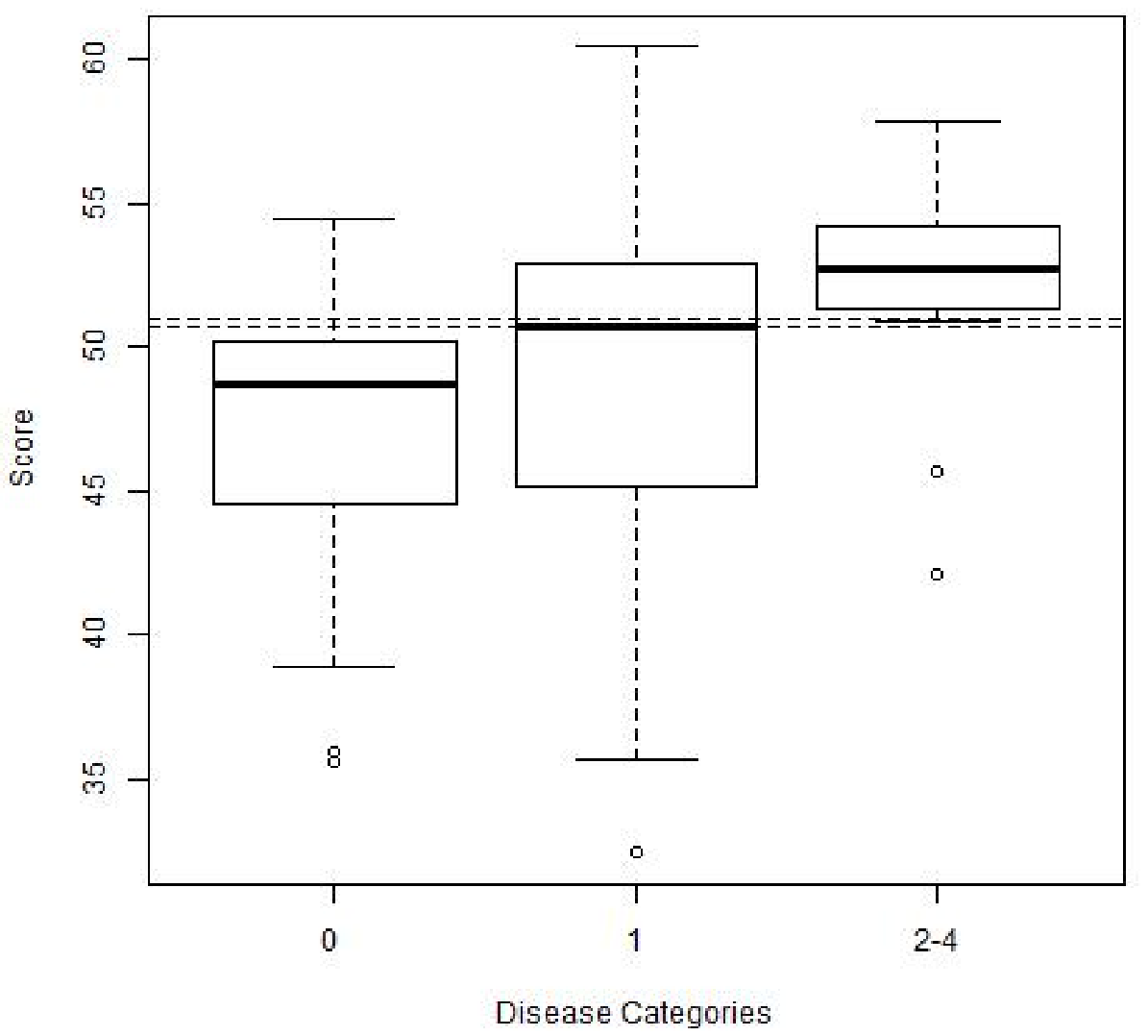}\label{fig:box04} }\\
	\subfigure[EHUM (SCOR) \newline $D_E = \mathbf{0.413}, YI = \mathbf{0.428}$] {\includegraphics[width=0.23\textwidth]{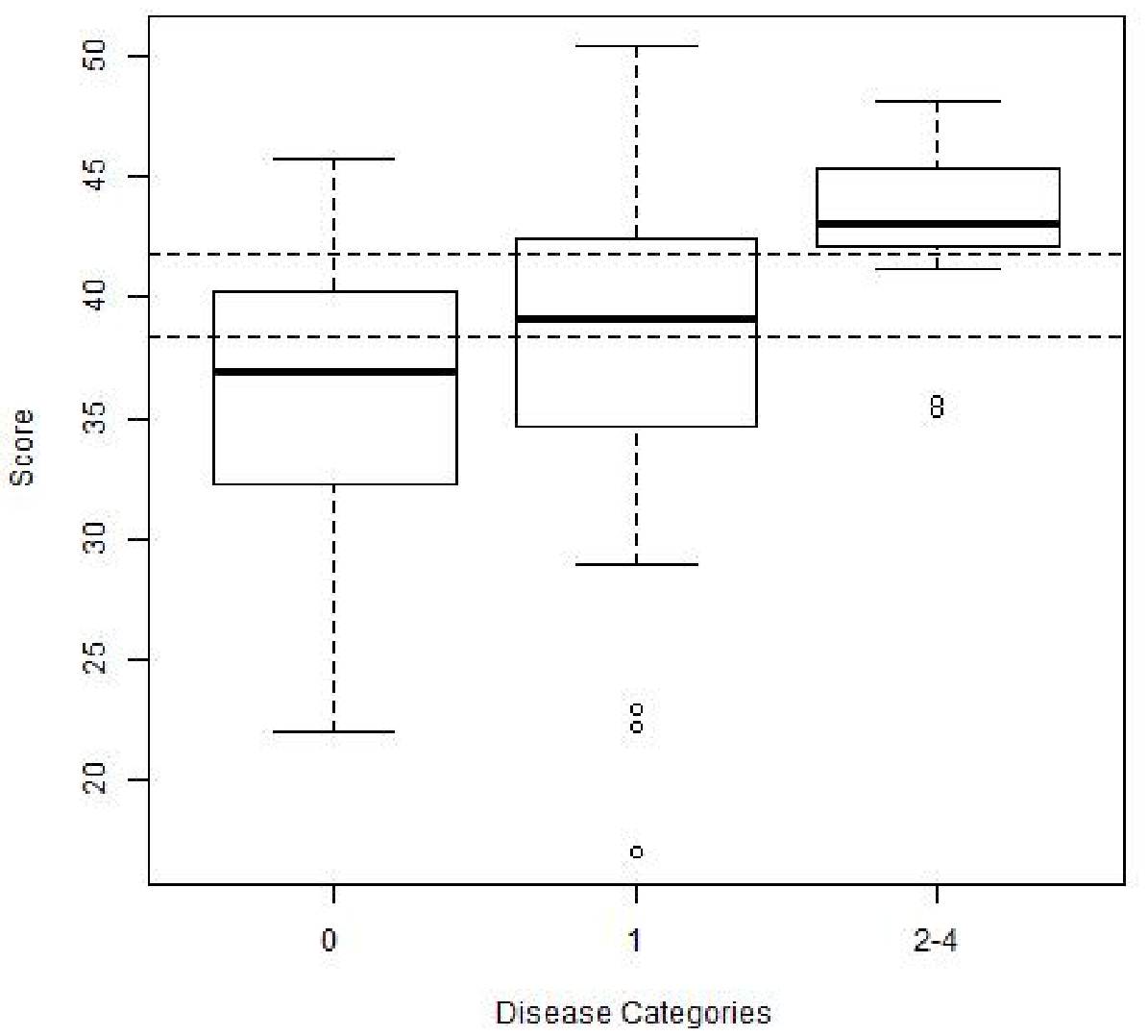}\label{fig:box05} }
	\subfigure[EHUM (NM) \newline $D_E = 0.341, YI = 0.334$] {\includegraphics[width=0.23\textwidth]{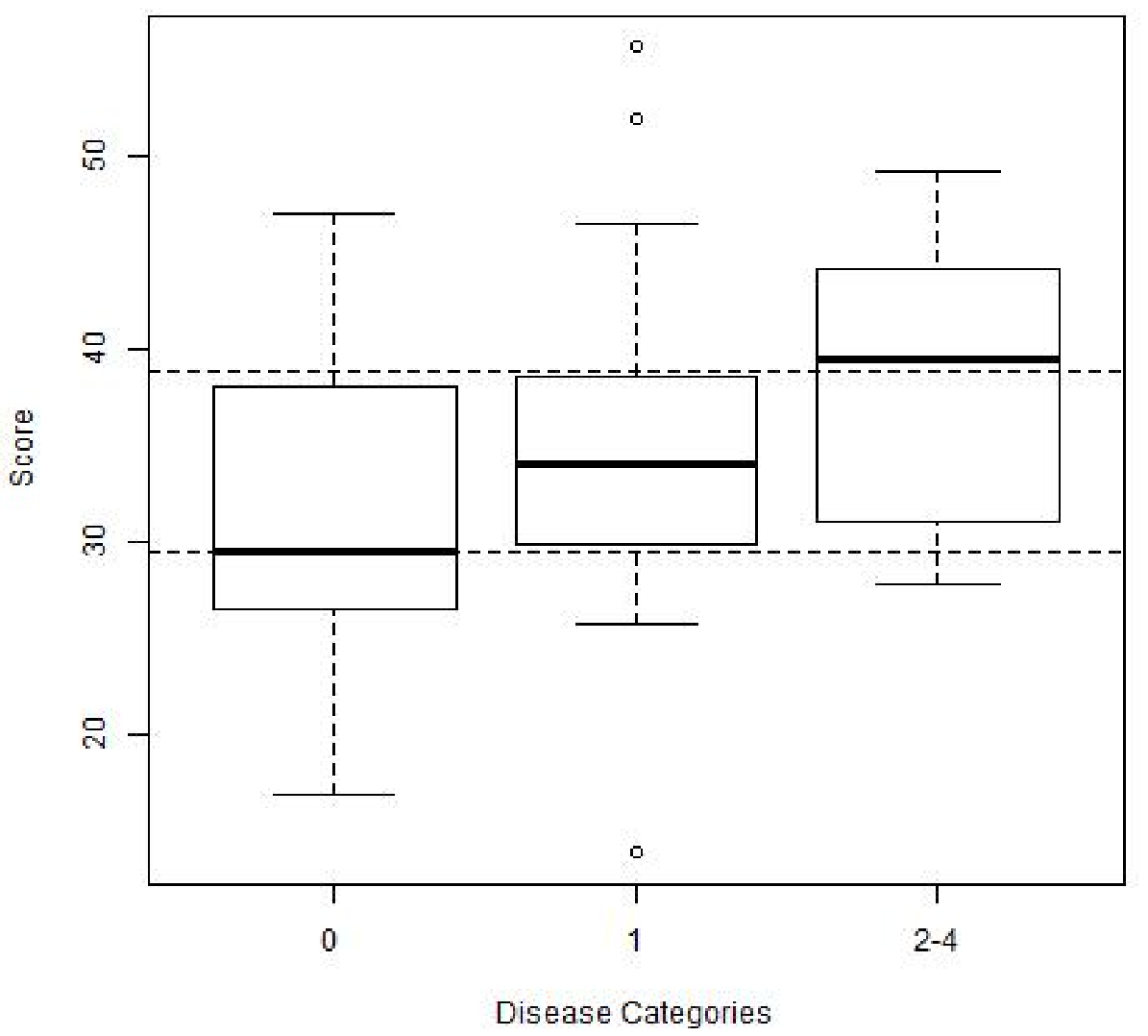}\label{fig:box06} }
	\subfigure[EHUM (Step-down) \newline $D_E = 0.341, YI = 0.408$] {\includegraphics[width=0.23\textwidth]{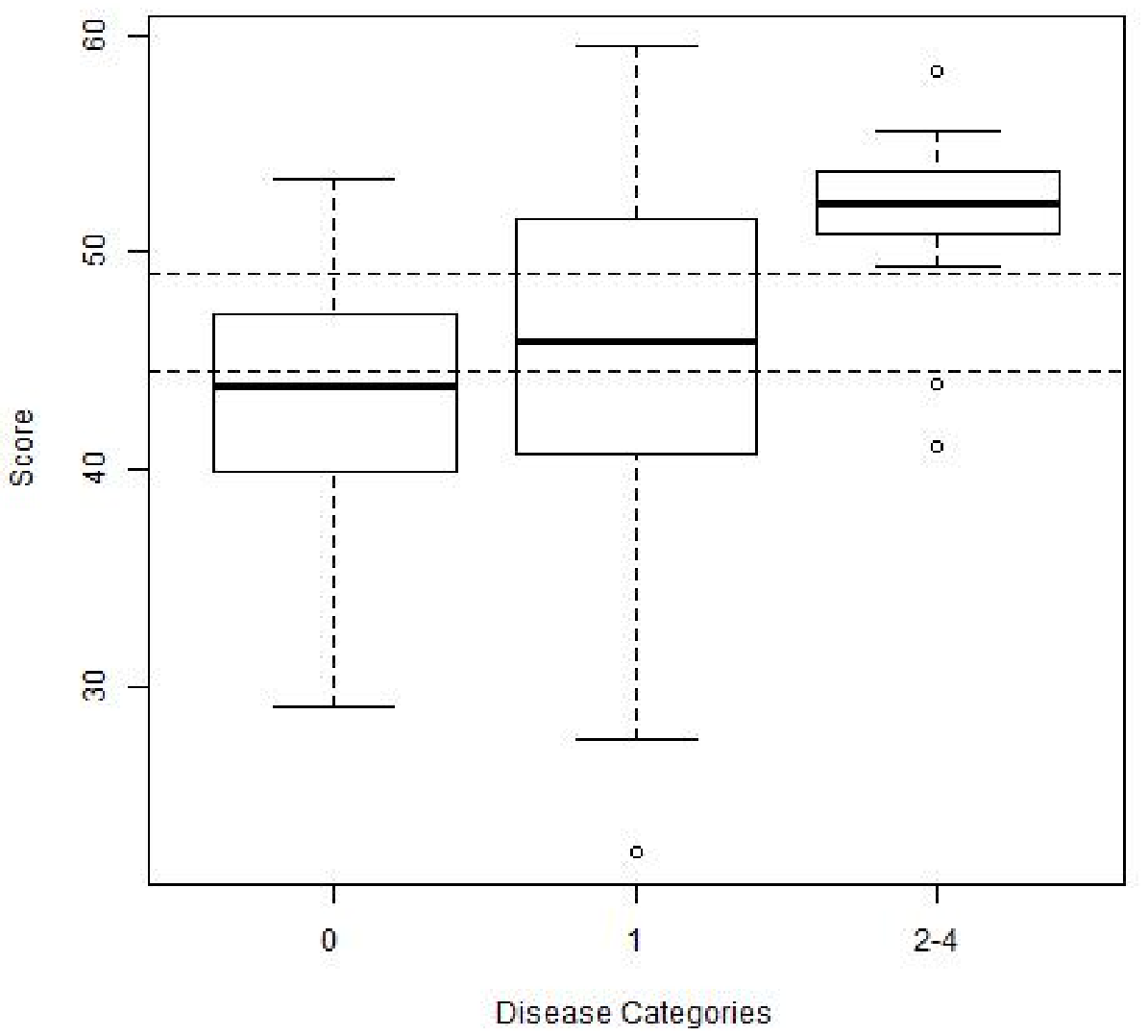}\label{fig:box07} }
	\subfigure[EHUM (Min-max) \newline $D_E = 0.363, YI = 0.361$] {\includegraphics[width=0.23\textwidth]{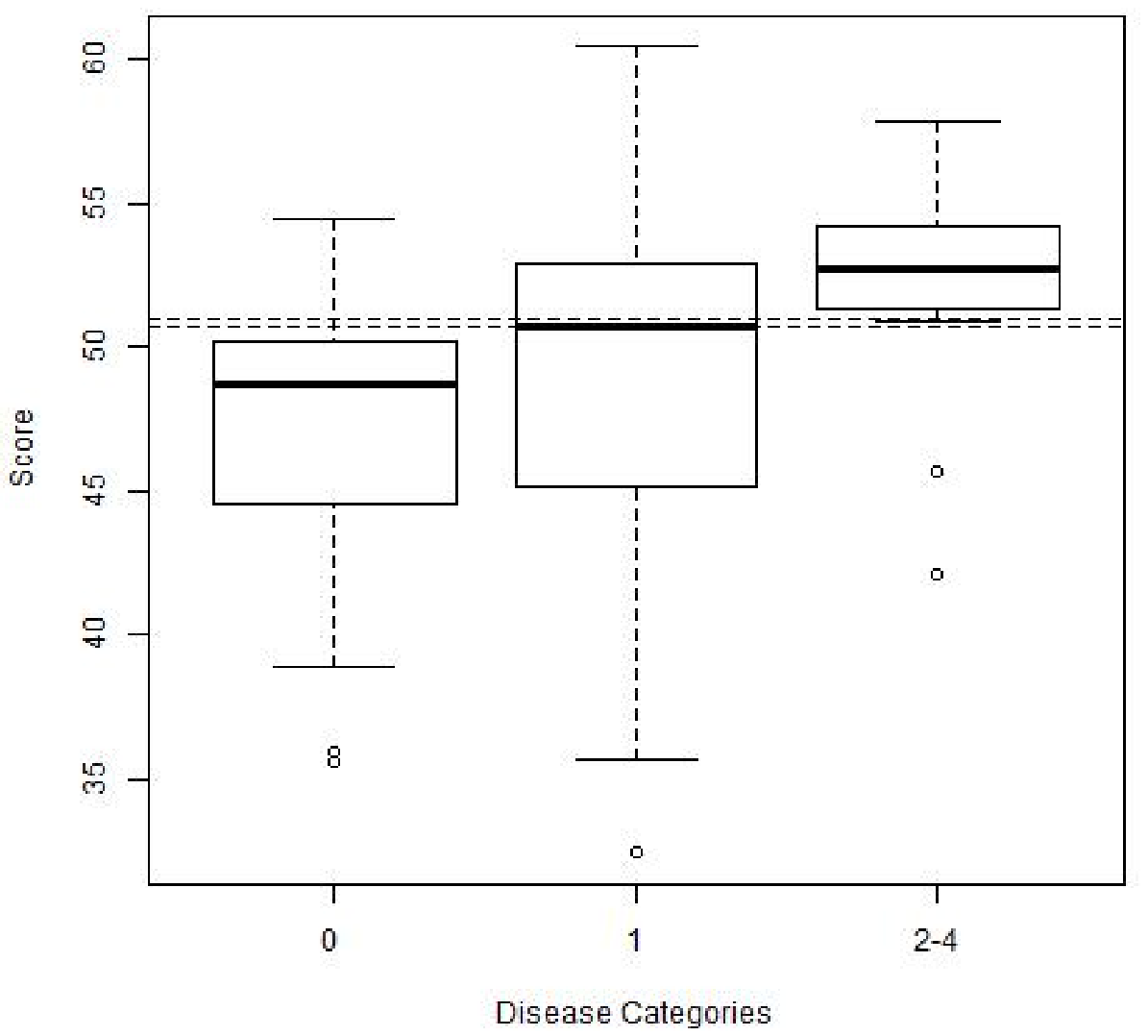}\label{fig:box08} }
\caption{Boxplots of the optimal combination vector scores are shown across the outcome categories 0, 1, and 2-4. The methods compared are SCOR ((a) ULBA, (e) EHUM), NM ((b) ULBA, (f) EHUM), Step-down ((c) ULBA, (g) EHUM) and Min-max ((d) ULBA, (h) EHUM). The horizontal dotted lines denote the corresponding cut-points for classification obtained by maximizing Youden's Index. Since this example includes 3 outcome categories, values of EHUM or ULBA greater than $\frac{1}{3\; !} \approx 0.167$ reflect improved accuracy over random guessing.}
\label{box_plots}
\end{figure}

Figure \ref{box_plots} shows the optimal combination vector scores for each class. SCOR results in improved separation of these scores across the classes, which is reflected in the higher values of EHUM and Youden's index achieved for both versions of the objective function  (ULBA and EHUM). In addition, the cut-points (marked by horizontal dotted lines) obtained by SCOR distinguish the three outcome categories more clearly  than those of step-down or min-max.

In Table \ref{bio_coeffs} we provide the values of the optimal combination coefficients using SCOR, Nelder-Mead, and step-down.  In the familiar setting of AUC estimation, where there are two outcome categories, a value of 0.5 corresponds to the expected value for random guessing, with higher values demonstrating improved classification accuracy.
Here, since we have three ordered categories, 
the expected value of the HUM under random guessing is $\frac{1}{3\; !} \approx 0.167$.
Hence, estimated values of EHUM or ULBA above 0.167 can be considered as an improvement over a random guess.
Since we obtain the highest value of $D_E(\hat{\beta})$ using the estimates obtained by maximizing ULBA with the SCOR algorithm, 
the coefficients obtained using this approach (i.e., those given in the first column) represent the preferred solution.
The esophagus and middle pharyngeal constrictor, which are known to be essential to swallowing function \citep{Levendag2007}, have the largest estimated coefficients, representing a relatively large contribution to the predicted combination scores. 
Since increasing radiation dose to structures in the head and neck is generally expected to worsen function, the slight negative coefficient for glottic area likely reflects correlation with other predictors, rather than a protective effect.
The results from this case study could be used to inform radiation treatment planning, since intensity-modulated radiation therapy plans could be adjusted to minimize predicted dysphagia risk by sparing radiation to key muscles and organs in the head-and-neck region.

\begin{table}[]
\centering
\resizebox{.9\columnwidth}{!}{%
\begin{tabular}{lcccccc}
\hline
Markers & ULBA (SCOR) & ULBA (NM) & ULBA (ST) & EHUM (SCOR) & EHUM (NM) & EHUM (ST) \\ \hline
Cricopharyngeal muscle & 0.243 & 0.703 & 0.137 & 0.293 & 0.697 & 0.137 \\
Esophagus & 0.518 & 0.293 & 0.302 & 0.600 & 0.299 & 0.302 \\
Glottic area & -0.080 & 0.534 & 0.061 & -0.100 & 0.539 & 0.061 \\
Inferior PC & 0.345 & 0.000 & 0.211 & 0.220 & 0.000 & 0.211 \\
Ipsilateral ADM & 0.311 & 0.216 & 0.092 & -0.009 & 0.215 & 0.092 \\
Middle PC & 0.509 & 0.102 & 0.698 & 0.570 & 0.097 & 0.698 \\
Mylohyoid muscle & 0.229 & 0.066 & 0.503 & 0.210 & 0.065 & 0.503 \\
Superior PC & 0.079 & 0.132 & 0.306 & 0.226 & 0.128 & 0.306 \\
Intrinsic tongue muscles & 0.365 & 0.237 & 0.000 & 0.274 & 0.239 & 0.000 \\ \hline
$D_E(\hat{\beta})$ & \textbf{0.419} & 0.341 & 0.399 & 0.413 & 0.341 & 0.399 \\ \hline
Youden Index & \textbf{0.471} & 0.334 & 0.408 & 0.428 & 0.334 & 0.408 \\ \hline
\end{tabular}}
\caption{Optimal coefficients obtained by maximizing the ULBA and EHUM objective functions using the SCOR, NM and step-down algorithms. The EHUM objective function values at all the obtained solutions and the estimated Youden's Index are reported in the last two rows.}
\label{bio_coeffs}
\end{table}

\vspace{-0.5cm}

\section{Discussion}
\label{sec_conc}
In this paper, we propose a novel derivative-free black-box optimization technique to minimize any non-convex function where the parameters are constrained to the surface of a unit sphere. 
The proposed algorithm is highly efficient, as it allows parallelization  using up to  $2n$ parallel threads when maximizing a function whose parameters belong to the surface of an $n$-dimensional unit sphere. Our simulations demonstrate that SCOR outperforms existing methods for biomarker combination, as well as other black-box optimization techniques, in terms of both performance and computation time. The R package \texttt{SCOR}, which implements the method described here, is available at \href{https://CRAN.R-project.org/package=SCOR}{https://CRAN.R-project.org/package=SCOR}.

We show that using SCOR, we obtain better estimates of the empirical hypervolume under the manifold (EHUM) compared to the estimates obtained  using the Nelder-Mead, step-down and min-max algorithms. Irrespective of the  objective function considered, the EHUM value at the  solution obtained by SCOR is always better than that obtained using the alternative algorithms. In our case study, we applied SCOR to find the optimal combination coefficients of radiation dose to various structures in the head-and-neck to distinguish subjects with increasing severity of swallowing difficulty experienced as a result of radiation treatment for cancer. We found that the values of EHUM and Youden's index obtained were higher for SCOR than for existing alternatives, suggesting that this approach can be applied to more accurately stratify patients based on their risk of developing dysphagia.


Here the proposed SCOR algorithm  is used mainly in the context of a classification problem with a hypervolume under manifolds criteria. However, this algorithm can be used in various other statistical problems such as directional statistics or single-index models where fixing the norm of the coefficient vector is needed to avoid the issue of non-identifiability. In future, the SCOR algorithms can be extended to the variable selection problem over the coefficients belonging to the surface of a unit sphere.


\backmatter


\section*{Acknowledgments}
We would like to acknowledge the following funding sources: grant MOE2015-T2-2-056 [Ministry of Education, Singapore], start-up grant [Duke-NUS Medical School, Singapore], cancer center support grant  P30CA016672 [National Institutes of Health/National Cancer Center], and  RP160668/RP150521 [Cancer Prevention and Research Institute of Texas].

\bigskip
\begin{center}
{\large\bf SUPPLEMENTARY MATERIAL}
\end{center}

\begin{description}

\item[Supplementary material:] The supplementary material file is provided in pdf format.

\item[R package:] The R package \texttt{SCOR}, available at \href{https://CRAN.R-project.org/package=SCOR }{https://CRAN.R-project.org/package=SCOR }, contains code to maximize the ULBA or EHUM objective functions for multi-category classification using the proposed method described in this article. This optimization method can also be used to maximize or minimize any black-box function on a spherically constrained parameter space.\\
The MATLAB code is  available at \href{https://github.com/priyamdas2/SCOR}{https://github.com/priyamdas2/SCOR}.


\end{description}
\vspace{-0.8cm}



\vspace{-.2cm}
\bibliographystyle{biom} 
\bibliography{biomsample}

\appendix




\label{lastpage}

\end{document}